\documentclass[twocolumn,aps,prl,preprintnumbers]{revtex4}
\usepackage{}
\usepackage{pdfpages}
\usepackage{float}
\usepackage{amsfonts}
\usepackage{bbm}
\usepackage[latin9]{inputenc}
\usepackage{amsmath}
\usepackage{amssymb}
\usepackage{graphicx}
\usepackage{mathrsfs}
\usepackage{amsfonts}
\usepackage{amsthm}
\usepackage{upgreek}
\usepackage{textcomp}
\usepackage{color}
\usepackage{txfonts}
\usepackage[colorlinks=true,citecolor=blue,linkcolor=blue,urlcolor=blue,anchorcolor=blue]{hyperref}%
\hypersetup{colorlinks=true,citecolor=blue,linkcolor=blue,urlcolor=blue}
\newcommand{\sign}[1]{\mathrm{sgn}(#1)}
\providecommand{\U}[1]{\protect\rule{.1in}{.1in}}
\makeatletter
\@ifundefined{textcolor}{}
{
\definecolor{BLACK}{gray}{0}
\definecolor{WHITE}{gray}{1}
\definecolor{RED}{rgb}{1,0,0}
\definecolor{GREEN}{rgb}{0,1,0}
\definecolor{BLUE}{rgb}{0,0,1}
\definecolor{CYAN}{cmyk}{1,0,0,0}
\definecolor{MAGENTA}{cmyk}{0,1,0,0}
\definecolor{YELLOW}{cmyk}{0,0,1,0}
}
\makeatother
\begin{document}

\title{Wave-propagation based analysis of the magnetostatic waves in ferrite films excited by metallic transducers}
\author{Zhizhi Zhang\textsuperscript{1}}
\author{Yuanming Lai\textsuperscript{1}}
\email[Corresponding author: ]{laiyuanming19@cdut.edu.cn}
\author{Qian Liu\textsuperscript{1}}
\author{Xiongzhang Liu\textsuperscript{1}}
\author{Chongsheng Wu\textsuperscript{1}}
\author{Peng Yan\textsuperscript{2}}
\email[Corresponding author: ]{yan@uestc.edu.cn}

\affiliation{1. School of Mechanical and Electrical Engineering, Chengdu University of Technology, Chengdu 610059, China}
\affiliation{2. School of Physics and State Key Laboratory of Electronic Thin Films and Integrated Devices, University of Electronic Science and Technology of China, Chengdu 611731, China}

\begin{abstract}
It is conventional wisdom that the spectra of the impedances of magnetostatic waves (MSWs) determine the transmissions of MSW devices. In the calculation of the impedances, microwave fields for MSW excitations are assumed to be uniform along the transducer and the metallicity of the transducers is seldom considered. In this work, a wave-propagation based analysis considering the nonuniform distributions of magnetic fields and the metallicity of the transducers is presented for investigating the propagations of MSWs and the performances of the device. Based on the proposed analyses, it is demonstrated that the nonreciprocities originate from the intrinsic asymmetries of the dispersion relations, the metallicity of the transducers causes the high insertion losses of the forward propagating waves in high-frequency bands, while the dips and severe in-band ripples in low-frequency bands are resulted from the complicated interference between the multiple width modes. The presented analyses are verified by the electromagnetic simulations and experiments with good agreements. Our work advances the understanding of MSWs propagating in ferrite films with metallic structures and paves the way to designing MSW devices aimed at the implantation in microwave systems.
\end{abstract}

\maketitle
\section{INTRODUCTION}

Magnetostatic waves (MSWs) are slowly propagating waves in ferro- or ferrimagnetic materials at frequencies up to several tens of GHz \cite{Chumak2015,Kruglyak2010,Serga2010,Lenk2011,Owens1985,Adam1991}. They are locally and collectively excited by radio frequency (RF) fields ($\mathbf{h}_{\mathrm{rf}}$s) alternatively applying torques on the statically biased magnetic moments under constant fields ($\mathbf{H}_0$s) \cite{Kasahara2017,Stancil2012,Kabos2012,Damon1961,Kalinikos1986}. Theories of MSWs have promoted the developments of many passive signal processing devices, such as magnetically tunable filters \cite{JWu2012,YZhang2020,XDu2024}, frequency selective limiters \cite{Adam2013,Geiler2021,HLin2023,Adam2004}, logic gates \cite{Jamali2013,Klingler2015,Kostylev2005}, delay lines \cite{Freer2019,Vysotski2006, Kobljanskyj2002,Bajpai1988,Wahi1982,Bajpai1989,Daniel1985}, isolators and circulators \cite{Szulc2020,Seewald2010,Shichi2015}, etc. In these devices, the most critical structures consist of two transducers on top of one magnetic film for the interconversions between MSWs and microwaves (MWs). The transducers take the forms of MW transmission lines, such as microstrips \cite{Thiancourt2024,Ganguly1975,Ganguly1978,Emtage1978,Sushruth2020}, co-plane waveguides \cite{Sushruth2020,YZhang2018,Ando1998,Birt2012,Celegato2015} or meander lines \cite{Nakrap2006,Collins1973,Sethares1979}, while the magnetic films are typically yttrium iron garnet (YIG), famous for its known smallest damping factor among all the magnetic materials \cite{Chang2014,Liu2014,Ding2020}.

This prerequisite has intrigued a plenty of studies on the mechanics and methods of ``MSW-MW" interconversions for high efficiencies and low insertion losses (ILs) \cite{Stancil2012,Kabos2012,Bajpai1988,Ganguly1975,Ganguly1978,Emtage1978,Ando1998,Sethares1979, Freire2003,Emtage1982,Aguilera2004,Kalinikos1981,Ando2003,Freire2003_IL,Sharawy1990,Lee1993,
Weiss2023Theo,Weiss2023Exp,Sethares1985,Chakrabarti1999,YLi1987,Vyzulin1993,Tsutsumi1996,Masuda1974, Tsutsumi1976}. Because the MSWs devices are mostly implanted in MW systems, conventional researches focus intensively on the calculation of the input and output impedances ($\mathbf{Z}_{\mathrm{in}}$s and $\mathbf{Z}_{\mathrm{out}}$s), to make the MSWs devices compatible. Toward this end, the transducers carrying the microwave surface current densities ($\mathbf{J}$s) are regarded as the sources of $\mathbf{h}_{\mathrm{rf}}$s. Through calculating the Poynting vectors, the power conveyed by MSWs can be expressed as the function of the $\mathbf{J}$s, where the $\mathbf{Z}_{\mathrm{in}}$s and $\mathbf{Z}_{\mathrm{out}}$s' dependencies on the magnetic parameters and the sizes of structures can be subsequently calculated \cite{Stancil2012,Kabos2012,Bajpai1988,Ganguly1975,Ganguly1978,Emtage1978,Ando1998,Sethares1979, Freire2003,Emtage1982,Aguilera2004,Kalinikos1981,Ando2003,Freire2003_IL,Sharawy1990,Lee1993,
Weiss2023Theo,Weiss2023Exp,Sethares1985,Chakrabarti1999,YLi1987}. These analyses succeeded to qualitatively predict the trend of the impedances' dependence on the electromagnetic parameters. However, they were not precise enough to obtain an acceptable approximation \cite{Ando2003}. Two possible reasons might be ascribed to: (1) the theoretical presumption of the uniform intensity of $\mathbf{h}_{\mathrm{rf}}$ along the magnetic film width may not be always valid \cite{Bajpai1988,Ganguly1975,Ganguly1978}; (2) the metallic nature of transducers is usually neglected \cite{Emtage1978,Sethares1979,Sethares1985}. Consequently, the underlying mechanism for the passband ripples and dips at low frequency bands and insertion losses at high frequency bands emerging in the transmission curves \cite{Serga2010,Freer2019,Bajpai1988,Geiler2022A,Geiler2022B} still remains to be fully understood.

In this paper, we theoretically investigate the propagation of MSWs in ferrite films excited by metallic transmission lines in a field analysis manner. The dispersion relations of MSWs are calculated, resulting in the nonreciprocal characteristics emerging in the ferrites with metallic layers. Specifically, the MSWs are unable to propagate through the ``metal/ferrite/metal" (MFM) structure, demonstrating that the metallic nature of the transducers causes the high ILs in high-frequency bands. Meanwhile, multiple width modes can be efficiently excited at some certain frequencies in low-frequency bands, whose interference might result in the destructive superpositions of converted MWs and the consequent dips and ripples in the transmission curves. Our analysis based on the propagating characteristics of MSWs in ferrite films provides more insight on the performance of the MSW devices, in sharp contrast to the traditional analysis focusing intensively on the characteristic impedances. Our work lays the foundation for fabricating MSW devices toward embedding into MW systems.

\section{ANALYTICAL MODEL}\label{Sec_AnaMod}
\begin{figure}[htbp!]
  \centering
  \includegraphics[width=0.48\textwidth]{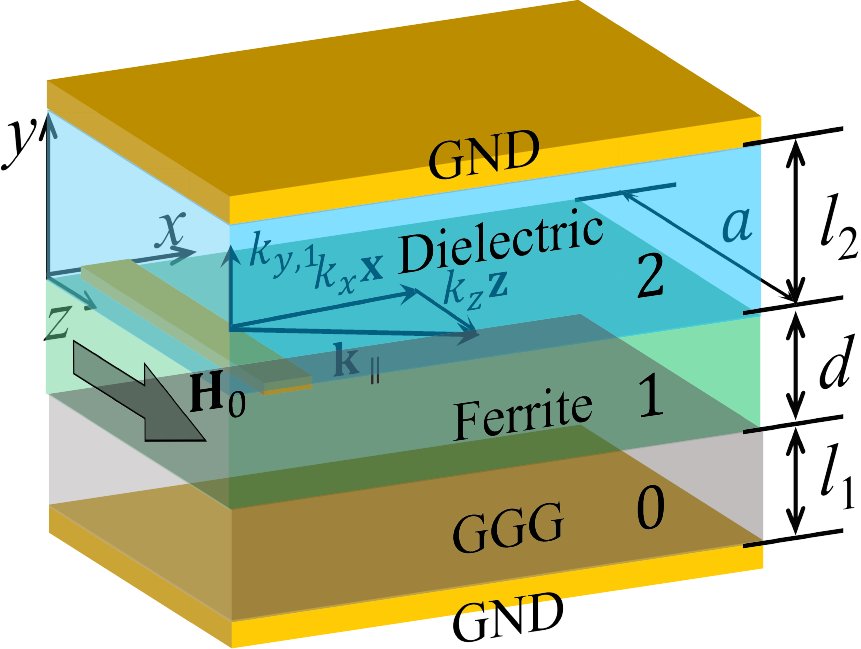}\\
  \caption{Schematic of the excitation and propagation of MSWs in the in-plane magnetized YIG layer. The MSWs, confined in the $y$ direction and indicated by the arrow labeled with $\mathbf{k}_{\parallel}$, are propagating in the $x-z$ plane.} \label{fig1}
\end{figure}

We consider a single ferrite (like YIG) film with the thickness $d$ and the width $a$ placed in the $x-z$ plane. It is extended along the $x$ direction and magnetized along the $z$ direction by the bias magnetic field $\mathbf{H}_\mathrm{0}=H_0{\mathbf{z}}$ (see Fig. \ref{fig1}). The YIG film is grown on the Gadolinium gallium garnet (GGG) substrate with thickness $l_1$, grounded by a perfect metallic layer (GND). The $\mathbf{h}_{\mathrm{rf}}$ for the MSW excitation is induced by the RF current flow through a microstrip line with length of $a$ on the top of the ferrite film, grounded by a substrate with thickness $l_2$. The top surface of the ferrite film is located at $y=0$.

The magnetization dynamics is governed by the Landau-Lifshitz-Gilbert equation:
\begin{equation} \label{LLG}
\frac{\partial\mathbf{M}_1}{\partial t}=-\gamma\mu_0\mathbf{M}_1\times\mathbf{H}_{\mathrm{eff,1}}+\frac{\alpha_1}{M_{\mathrm{s,1}}}\mathbf{M}_1\times\frac{\partial\mathbf{M}_1}{\partial t},
\end{equation}
where $\gamma$ is the gyromagnetic ratio; $\mu_0$ is the vacuum permeability; $\alpha_1\ll1$ is the dimensionless Gilbert damping constant; $M_{\mathrm{s,}1}$ is the saturated magnetization; $\mathbf{M}_1$ is the magnetization and $\mathbf{H}_{\mathrm{eff,}1}$ is the effective magnetic field. $\mathbf{H}_{\mathrm{eff,}1}$ comprises $\mathbf{H}_\mathrm{0}$, the intralayer exchange field $\mathbf{H}_{\mathrm{ex},1}$, the dipolar field $\mathbf{H}_{\mathrm{d},1}$ and the $\mathbf{h}_{\mathrm{rf}}$.

We adopt the linear approximation, which is $\mathbf{M}_1=\mathbf{M}_{d,1}+M_{\mathrm{s,}1}\mathbf{z}$, with $\mathbf{M}_{d,1}=M_{x\mathrm{,}1}\mathbf{x}+M_{y\mathrm{,1}}\mathbf{y}$ being the dynamic component and possessing the form $\mathbf{M}_{d,1}=\mathbf{M}_{01}e^{i\left(\omega t-\mathbf{k}_1{\cdot}\mathbf{r}\right)}$, where $\mathbf{M}_{01}=M_{0x\mathrm{,}1}\mathbf{x}+M_{0y\mathrm{,}1}\mathbf{y}, M_{0x\mathrm{,}1},\ M_{0y\mathrm{,}1}\ll M_{\mathrm{s,}1}$, $\omega$ being the angular frequency of spin precession and $\mathbf{k}_1=\mathbf{k}_{\parallel,1}+k_{y,1}\mathbf{y}$ with $\mathbf{k}_{\parallel,1}=k_{x,1}\mathbf{x}+k_{z,1}\mathbf{z}$ being the in-plane wave vectors. We note that $k_{x,1}$, $k_{y,1}$ and $k_{z,1}$ can be either real or purely imaginary, indicating that the waves are either propagating or evanescent in the corresponding directions \cite{Damon1961}. The $\mathbf{H}_{\mathrm{ex},1}$ is given by:
\begin{equation} \label{H_ex}
\begin{aligned}
\mathbf{H}_{\mathrm{ex},1}=\frac{2A_{\mathrm{ex},1}}{\mu_0M_{\mathrm{s,1}}^2}\nabla^2\mathbf{M}_1 =l_{\mathrm{ex},1}^2\nabla^2\mathbf{M}_1,
\end{aligned}
\end{equation}
with $A_{\mathrm{ex},1}$ being the exchange constant and $l_{\mathrm{ex},1}$ being the exchange length \cite{Abo2013}, which is about 16 nm by the magnetic parameters of YIG. Therefore, we neglect $\mathbf{H}_{\mathrm{ex},1}$ when the wavelength is in the range of several tens of micrometers. The MSWs in the magnetic layers satisfy the magnetostatic limit, such that the Maxwell equations describing the electrodynamics in the whole regions are:
\begin{subequations}  \label{MagnStatMAX}
\begin{align}
\label{MagnStatMAXa} \nabla\cdot\mathbf{B}_q= \mu_0\nabla\cdot\left(\mathbf{H}_{\mathrm{d},q}+\mathbf{M}_q\right)=0, \\
\label{MagnStatMAXb} \nabla\times\mathbf{H}_{\mathrm{d},q}=0,  &
\end{align}
\end{subequations}
where the subscript q=0,1,2 labels the regions shown in Fig. \ref{fig1}, respectively.
\section{THEORIES}\label{Sec_Theo}

\subsection{Formulation}\label{Formu}

Considering the curveless nature of $\mathbf{H}_{\mathrm{d},q}$ described by Eq. (\ref{MagnStatMAXb}), we can define a magnetic scale potential $\varphi_q$ satisfying $\mathbf{H}_{\mathrm{d},q}=-\nabla\varphi_q$. In the MSW dispersion relation analysis, we firstly neglect $\mathbf{h}_{\mathrm{rf}}$ considering that MSWs propagates far away from microstrip lines, then LLG equation reads:

\begin{subequations}  \label{LLGExpd}
\begin{align}
\label{LLGExpda} i\omega M_{x\mathrm{,1}}=-\omega_HM_{y\mathrm{,1}}-\omega_{M,1}\frac{\partial\varphi_1}{\partial y}-i\omega\alpha_1M_{y\mathrm{,1}} \\
\label{LLGExpdb} i\omega M_{y\mathrm{,1}}=\omega_HM_{x\mathrm{,1}}+\omega_{M,1}\frac{\partial\varphi_1}{\partial y}+i\omega\alpha_1M_{x\mathrm{,1}} &
\end{align}
\end{subequations}
where $i$ is the imaginary unit, $\omega_{M,1}=\gamma\mu_0M_{\mathrm{s,}1}$ and $\omega_H=\gamma\mu_0H_0$. Therefore, the relation between $\mathbf{M}_{d,1}$ and $\varphi_1$ can be derived from Eqs. (\ref{LLGExpd}) that

\begin{equation} \label{M_dipFld}
\begin{aligned}
\left[\begin{matrix}M_{x\mathrm{,1}}\\M_{y\mathrm{,1}}\\\end{matrix}\right] = \overrightarrow{\mathbf{\chi}}\left[\begin{matrix}-\frac{\partial\varphi_1}{\partial x}\\-\frac{\partial\varphi_1}{\partial y}\\\end{matrix}\right],
\end{aligned}
\end{equation}
with $\overrightarrow{\mathbf{\chi}}$ being the susceptibility tensor defined as \cite{Polder1949}:
\begin{equation} \label{Sus_Tensor}
\begin{aligned}
\overrightarrow{\mathbf{\chi}}= \left[\begin{matrix}\chi_1&i\kappa_1\\-i\kappa_1&\chi_1\\\end{matrix}\right],
\end{aligned}
\end{equation}
where
\begin{subequations}  \label{Sus_Comp}
\begin{align}
\label{Sus_Compa} \chi_1=\frac{\left(\omega_H+i\alpha_1\omega\right)\omega_{M,1}} {\left(\omega_H+i\alpha_1\omega\right)^2-\omega^2}, \\
\label{Sus_Compb} \kappa_1=\frac{\omega\omega_{M,1}} {\left(\omega_H+i\alpha_1\omega\right)^2-\omega^2}.  &
\end{align}
\end{subequations}
Substituting Eqs. (\ref{Sus_Tensor}) and (\ref{Sus_Comp}) into Eq. (\ref{LLGExpda}), we obtain:
\begin{subequations}  \label{WalkerEqs}
\begin{align}
\label{WalkerEqsa}\frac{\partial^2\varphi_q}{\partial x^2}+\frac{\partial^2\varphi_q}{\partial y^2}+\frac{\partial^2\varphi_q}{\partial z^2}=0,\ q=0,2 \\
\label{WalkerEqsb}\left(1+\chi_1\right)\left(\frac{\partial^2\varphi_1}{\partial x^2}+\frac{\partial^2\varphi_1}{\partial y^2}\right)+\frac{\partial^2\varphi_1}{\partial z^2}=0.  &
\end{align}
\end{subequations}
We consider that $\varphi_q$s should take the same harmonic terms with $\mathbf{M}_1$, hence:
\begin{subequations}  \label{WalkerEqsVec}
\begin{align}
\label{WalkerEqsVeca}k_{x,q}^2+k_{y,q}^2+k_{z,q}^2=0,\ q=0,2 \\
\label{WalkerEqsVecb}\left(1+\chi_1\right)\left(k_{x,1}^2+k_{y,1}^2\right)+k_{z,1}^2=0.  &
\end{align}
\end{subequations}
The $k_{y,1}$ has two possible solutions with opposite signs. The $\varphi_q$s is supposed to take the following forms:
\begin{subequations}  \label{PotentialExp}
\begin{align}
\label{PotentialExpa}\varphi_q=& \left(A_qe^{k_{\parallel,q}y}+C_qe^{-k_{\parallel,q}y}\right)e^{i\left(\omega t-\mathbf{k}_{\parallel,q}\cdot\mathbf{r}_\parallel\right)},\ q=0,2,\\
\label{PotentialExpb}\varphi_1=& \left[A_1\sin{\left(k_{y,1}y\right)}+C_1\cos{\left(k_{y,1}y\right)}\right]e^{i\left(\omega t-\mathbf{k}_{\parallel,1}\cdot\mathbf{r}_\parallel\right)},
\end{align}
\end{subequations}
where $k_{\parallel,q}^2=k_{x,q}^2+k_{z,q}^2$. Here, we note that $k_{x,q}$ and $k_{z,q}$ can only be real since MSWs are propagating in the $x-z$ plane. Therefore, the in-plane components of $\mathbf{H}_{\mathrm{d},q}$ can be expressed as:
\begin{subequations}  \label{H_paraExp}
\begin{align}
\label{H_paraExpa}\mathbf{H}_{\mathrm{{\parallel}d},q}&=i\mathbf{k}_{\parallel,q} \left(A_qe^{k_{\parallel,q}y}+C_qe^{-k_{\parallel,q}y}\right)e^{i\left(\omega t-\mathbf{k}_{\parallel,q}\cdot\mathbf{r}_\parallel\right)},\ q=0,2,\\
\label{H_paraExpb}\mathbf{H}_{\mathrm{{\parallel}d},1}&=i\mathbf{k}_{\parallel,1} \left[A_1\sin{\left(k_{y,1}y\right)}+C_1\cos{\left(k_{y,1}y\right)}\right]e^{i\left(\omega t-\mathbf{k}_{\parallel,1}\cdot\mathbf{r}_\parallel\right)}.
\end{align}
\end{subequations}
In consideration of the continuity of $\mathbf{H}_{\mathrm{{\parallel}d},q}$ at the interfaces ($y=-d_1\mathrm{\ \mathrm{and}\ }0$, respectively), it can be concluded that the $k_{\parallel,q}$s and $k_{\parallel,qj}$s are identical. For simplification, let's denote them as $\mathbf{k}_\parallel=k_x\mathbf{x}+k_z\mathbf{z}$ with $k_\parallel$ being the modulus. Then Eqs. (\ref{H_paraExp}) can be rewritten as:
\begin{subequations}  \label{H_paraExpkPara}
\begin{align}
\label{H_paraExpkParaa}\mathbf{H}_{\mathrm{{\parallel}d},q}=& i\mathbf{k}_\parallel\left(A_qe^{k_\parallel y}+C_qe^{-k_\parallel y}\right)e^{i\left(\omega t-\mathbf{k}_\parallel\cdot\mathbf{r}_\parallel\right)},\ q=0,2,\\
\label{H_paraExpkParab}\mathbf{H}_{\mathrm{{\parallel}d},1}=& i\mathbf{k}_\parallel\left[A_1\sin{\left(k_{y,1}y\right)}+C_1\cos{\left(k_{y,1}y\right)}\right]e^{i\left(\omega t-\mathbf{k}_\parallel\cdot\mathbf{r}_\parallel\right)}.
\end{align}
\end{subequations}
The out-of-plane components of $\mathbf{H}_{\mathrm{d},1}$ can be expressed as:
\begin{equation} \label{H_Perp}
\begin{aligned}
H_{y\mathrm{d},1}= k_\parallel\left[-A_1\cos{\left(k_{y,1}y\right)}+ C_1\sin{\left(k_{y,1}y\right)}\right]e^{i\left(\omega t-\mathbf{k}_\parallel\cdot\mathbf{r}_\parallel\right)}.
\end{aligned}
\end{equation}
The out-of-plane component of $\mathbf{B}_q$ can be expressed as:
\begin{widetext}
\begin{subequations}  \label{B_Perp}
\begin{align}
\label{B_Perpa}B_{y,q}&=-\mu_0k_\parallel\left(A_qe^{k_\parallel y}-C_qe^{-k_\parallel y}\right)e^{i\left(\omega t-\mathbf{k}_\parallel\cdot\mathbf{r}_\parallel\right)},\ q=0,2,\\
\label{B_Perpb}B_{y,1}&=\mu_0\left\{\left[\kappa_1\sin{\left(k_{y,1}y\right)}k_x- \left(\chi_1+1\right)\cos{\left(k_{y,1}y\right)}k_{y,1}\right]A_1+
\left[\kappa_1\cos{\left(k_{y,1}y\right)}k_x+ \left(\chi_1+1\right)\sin{\left(k_{y,1}y\right)}k_{y,1}\right]C_1\right\}e^{i\left(\omega t-\mathbf{k}_\parallel\cdot\mathbf{r}_\parallel\right)}.
\end{align}
\end{subequations}
\end{widetext}

The boundary conditions are the $\mathbf{H}_{\mathrm{{\parallel}d},q}$s' continuity at the interfaces, and the $B_{y,q}$s' continuity at the interfaces and their vanishment at metallic boundaries ($y=-d-l_1\ \mathrm{and\ }l_2$, respectively). These conditions give the following equations:

\begin{widetext}
\begin{subequations}  \label{BondCond}
\begin{align}
\label{BondConda}A_0e^{-k_\parallel\left(d+l_1\right)}-C_0e^{k_\parallel\left(d+l_1\right)}&=0,\\
\label{BondCondb}A_0e^{-k_\parallel d}+C_0e^{k_\parallel d}&=-A_1\sin{\left(k_{y,1}d\right)}+C_1\cos{\left(k_{y,1}d\right)},\\
    \label{BondCondc}-k_\parallel\left(A_0e^{-k_\parallel d}-C_0e^{k_\parallel d}\right)&=\left[-\kappa_1\sin{\left(k_{y,1}d\right)}k_x- \left(\chi_1+1\right)\cos{\left(k_{y,1}d\right)}k_{y,1}\right]A_1
    + \left[\kappa_1\cos{\left(k_{y,1}d\right)}k_x- \left(\chi_1+1\right)\sin{\left(k_{y,1}d\right)}k_{y,1}\right]C_1,\\
\label{BondCondd}C_1&=A_2+C_2,\\
\label{BondConde}-k_\parallel\left(A_2-C_2\right)&=\left[-\left(\chi_1+1\right)k_{y,1}\right]A_1+ \left(\kappa_1k_x\right)C_1,\\
\label{BondCondf}A_2e^{k_\parallel l_2}-C_2e^{-k_\parallel l_2}&=0.
\end{align}
\end{subequations}
\end{widetext}
Equations (\ref{BondCond}) are linear and homogenous for the variables $A_q$s and $C_q$s. The nonzero solutions require a vanishing determinant of the coefficient matrix, which gives \cite{Yukawa1977}:
\begin{widetext}
\begin{equation} \label{DispRela}
\begin{aligned}
\left[k_\parallel\tanh{\left(k_\parallel l_2\right)}-k_x\kappa_1\right]\left[k_\parallel\tanh{\left(k_\parallel l_1\right)}+k_x\kappa_1\right]+\left(\chi_1+1\right)k_{y,1}\cot{\left(k_{y,1}d\right)}k_\parallel\left[\tanh{\left(k_\parallel l_1\right)}+\tanh{\left(k_\parallel l_2\right)}\right]-k_{y,1}^2\left(\chi_1+1\right)^2=0.
\end{aligned}
\end{equation}
\end{widetext}

The dependence of $\omega$ on the combination of $k_x$ and $k_z$ can then be determined by solving the combination of Eqs. (\ref{WalkerEqsVecb}) and (\ref{DispRela}), which give the dispersion relation of MSWs in ferrite films. We note that Eq. (\ref{DispRela}) is transcendental, so the dispersion relation can be typically calculated by graphic methods, which heavily rely on the programming techniques and the capability of computers for acceptable accuracies.

\subsection{Analytical calculation}\label{Ana_cal}

We consider a simple case that the $\mathbf{h}_{\mathrm{rf}}$ is uniform along the ferrite width, so that the excited MSWs take $k_z=0$. This can be reasonable when the microstrip line is shorted at the end and its portion on the ferrite film is an appreciable fraction of the MW wavelength at low frequencies. The state-of-the-art micro-fabrication enables the microstrip lines being directly deposited on the ferrite, corresponding to $l_2\rightarrow+\infty$. A single-crystal YIG film with $\alpha_1\approx0$ can be liquid phase epitaxially grown to the required thickness. The GGG substrates can also be polished to the designed thickness or even perfectly removed. Therefore, the $d$ and $l_1$ can be variables.

Substituting $k_z=0$ into Eqs. (\ref{WalkerEqsVecb}) and (\ref{DispRela}), we obtain:
\begin{equation} \label{Eq29}
\begin{aligned}
k_{y,1}=\pm ik_x,
\end{aligned}
\end{equation}
\begin{equation} \label{Eq30}
\begin{aligned}
\left[\sign{k_x} - \kappa_1\right] & \left[\tanh{\left(k_xl_1\right)}+ \kappa_1\right]+\left(\chi_1+1\right)^2\\  + \left(\chi_1+1\right) & \coth{\left(k_xd\right)}  \left[\tanh{\left(k_xl_1\right)}+\sign{k_x}\right]=0,
\end{aligned}
\end{equation}
where the identical transformations $\cot{\left(ix\right)}=-i\coth{\left(x\right)}$ and $\tanh{\left(k_xl_2\right)}=\sign{k_x}$ with $l_2\rightarrow+\infty$ are adopted.

\begin{figure}[htbp!]
  \centering
  \includegraphics[width=0.48\textwidth]{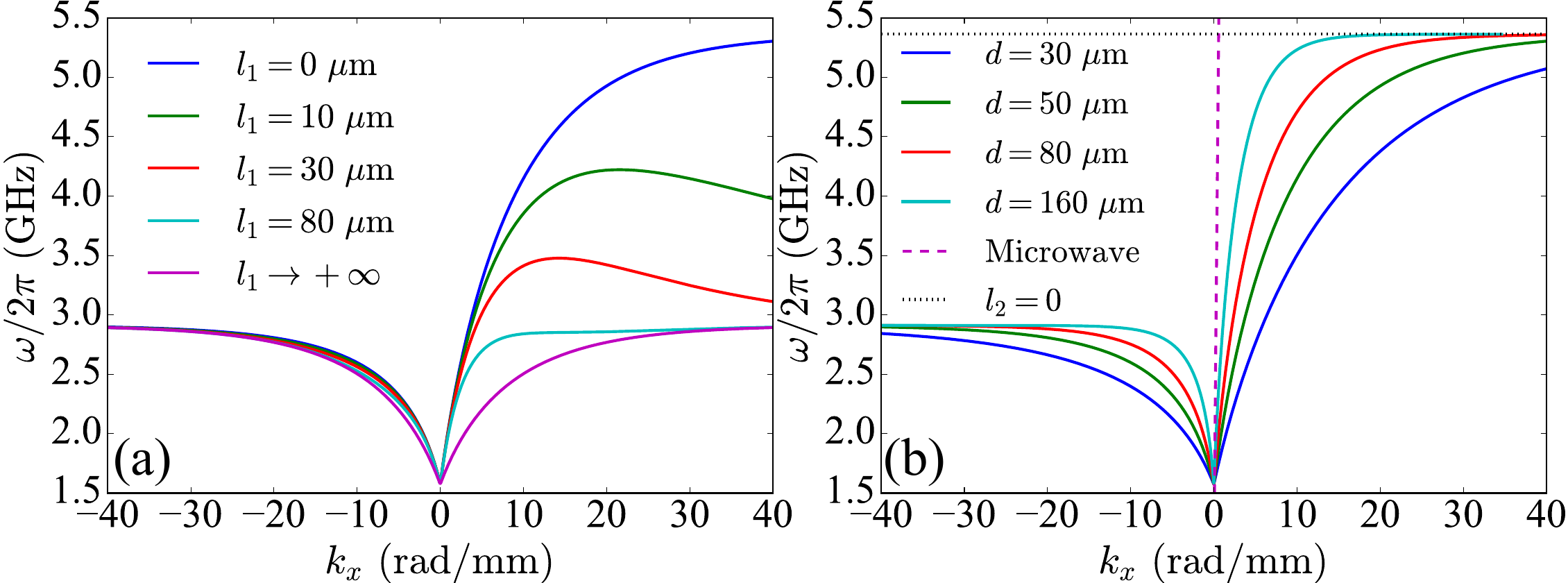}\\
  \caption{Numerical results of the MSWs dispersion curves with (a) $d=50\mathrm{\ \mu m}$ but varied $l_1$ ranging from 0 to $+\infty$ and (b) $l_1=0$ but varied $d$ ranging from         $30\ \mathrm{\mu m}$ to $160\ \mathrm{\mu m}$. The dashed magenta and the dotted black lines in (b) represent the dispersion relations of MWs and MSWs with $l_2=0$, respectively.} \label{fig2}
\end{figure}

The $M_{\mathrm{s,}1}$ of YIG is $1.48\times{10}^5\ \mathrm{A/m}$. In the following calculation, we set $H_0=16.5\ \mathrm{mT}$ if not stated otherwise. Firstly, we fixed $d=50\ \mathrm{\mu m}$. The MSWs dispersion relation curves with $l_1$ ranging from 0 to $+\infty$ are plotted in Fig. \ref{fig2}(a), coinciding with the results by T. Yukawa, et. al. \cite{Yukawa1977}. The dispersion relations have the following features.

(i) The $l_1$ has negligible impacts on the dispersion relations when $l_1\gg d$, implying the metallic GND layers cannot influence the propagation of MSWs if they are far away from the ferrite films.

(ii) The $k_{y,1}$s are always purely imaginary, demonstrating that the intensities of dynamic magnetizations are attenuating along the thickness of films. The energies of such MSWs are localized at the surfaces, so they are nominated as surface MSWs, while those taking real $k_{y,1}$ values are nominated as volume MSWs \cite{Damon1961}.

(iii) The slope of the dispersion curve for forward MSWs (whose $k_x>0$)  increases with the $l_1$ decreasing, while that for backward MSWs (whose $k_x<0$) remains almost the same. Here, the nonreciprocities emerge with the GND layer getting closer to one surface of the ferrite films.

(iv) When the $l_1$ is shorter than a certain distance, which is $80\ \mathrm{\mu m}$ in our case, the dispersion curve for forward MSWs shall reach a maximal value, and then decline with the $k_x$ increasing, as shown by the red and green curves in  Fig. 2(a).  It means that in the structures with such $l_1$s, excited MSWs might have positive phase velocities ($v_p$s) but negative group velocities ($v_g$s), considering $v_p={\omega}/{k_x}$ and $v_g={\partial\omega}/{\partial k_x}$.

(v) When $l_1=0$, together with $l_2\rightarrow+\infty$ forming the ``ferrite/metal" (FM) structure, the frequency of forward MSWs keeps on rising and finally saturates with the $k_x$ increasing. It is found that the forward MSWs cover the widest frequency band in FM structures.

In the following analysis, we set $l_1=0$ for the widest frequency band of forward MSWs and the positive $v_g$ in the full frequency bands. The MSWs dispersion relation curves with d ranging from $30\ \mathrm{\mu m}$ to $160\ \mathrm{\mu m}$ are plotted as the solid lines in Fig. \ref{fig2} (b). The MW's dispersion relation is calculated by the formula $\omega={k_x}/{\sqrt{\epsilon\mu_0}}$ with $\epsilon$ being the permittivity of YIG, plotted as the dashed magenta line in Fig. \ref{fig2} (b) for a comparison. We notice: (i) the slopes of dispersion curves for MSWs are much lower than their MW counterpart, reconfirming that MSWs are slow waves. (ii) The MSW frequency starts at $\omega_{\mathrm{low}}=\sqrt{\omega_H\left(\omega_H+\omega_{M,1}\right)}$ with zero $k_x$, and saturates at $\omega_{\mathrm{cut,\ b}}=\omega_H+0.5\omega_{M,1}$  and $\omega_{\mathrm{cut,\ f}}=\omega_H+\omega_{M,1}$ for the backward and forward MSWs, respectively. (iii) The MSWs dispersion curve gets closer to the MW's with $d$ increasing. The increasing slope manifests the $v_g$s and the wavelengths of MSWs at the same frequencies are increasing. Especially, we are also interested in the case that both $l_1$ and $l_2$ are zero, forming the MFM structure. By substituting $l_2=0$ into Eqs. (\ref{Eq29}) and (\ref{Eq30}), we find that $\omega=\omega_H+\omega_{M,1}$ is a constant independent of $k_x$, indicating that the dispersion curve is a horizontal line, plotted as the dotted black line in Fig. \ref{fig2}(b). It means the MFM structures do not support the MSWs propagating along $x$ direction.

\begin{figure}[htbp!]
  \centering
  \includegraphics[width=0.48\textwidth]{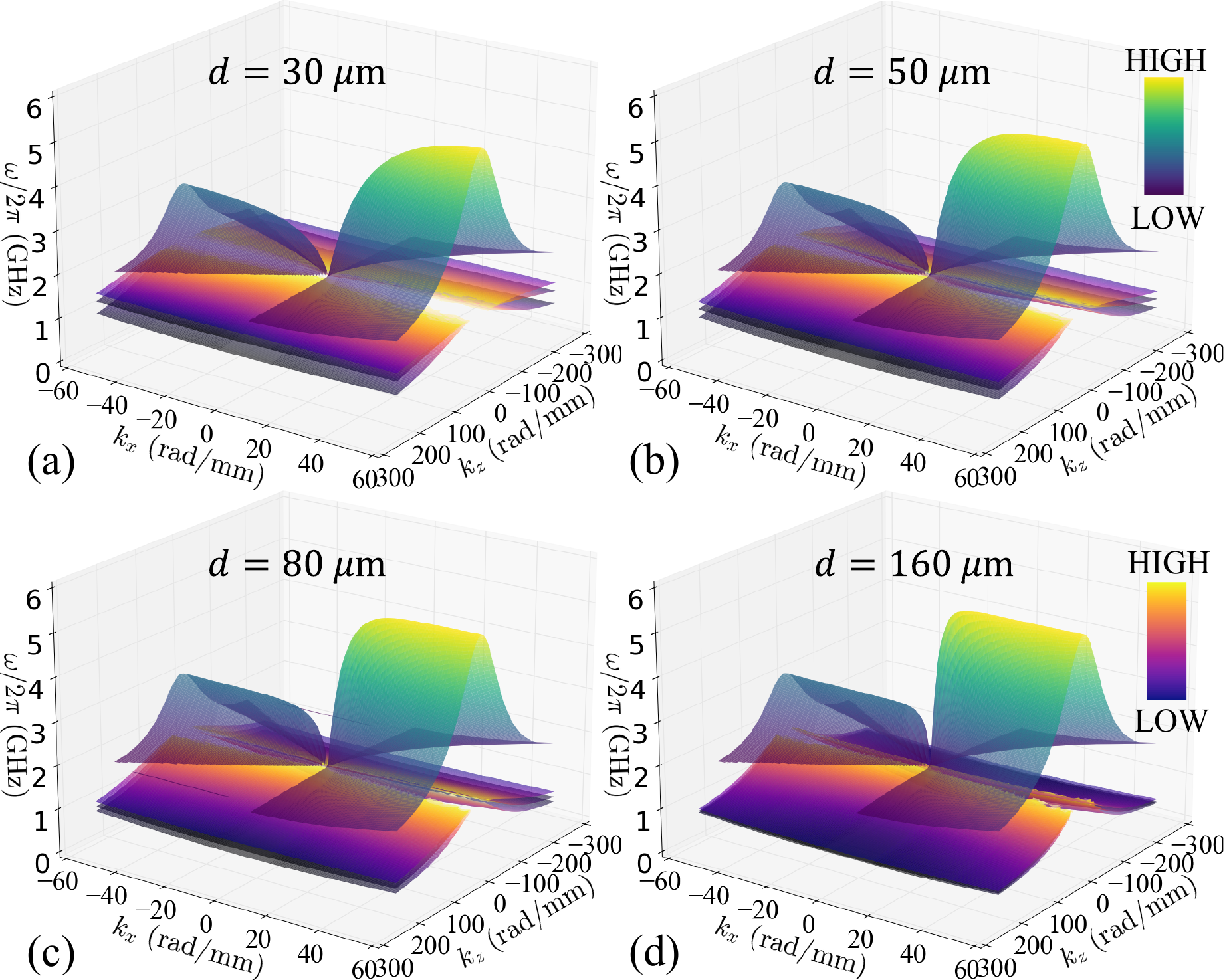}\\
  \caption{Numerical results of the MSWs dispersion surfaces with $l_1=0$ but different (a) $d=30\mathrm{\ \mu m}$, (b) $d=50\mathrm{\ \mu m}$, (c) $d=80\mathrm{\ \mu m}$ and (d) $d=160\mathrm{\ \mu m}$. The surfaces rendered by viridian and plasma colormaps represent the solutions of surface and volume modes, respectively. The color bars in (b) and (d) interpret the frequencies in surface and volume modes, respectively.} \label{fig3}
\end{figure}

Next, we investigate more general cases with $k_z\neq0$, in which the variables $k_x$ and $k_z$ jointly determine the $\omega$. Then the dispersion curves shall evolve into dispersion surfaces, as plotted in Fig. \ref{fig3}. The surfaces rendered by the viridian and plasma colormaps represent the solutions with $k_{y,1}$ being purely imaginary and real, corresponding to the surface and volume modes, respectively. A single surface MSW mode and multiple volume MSW modes exist in the ferrite films. Here, only the first three volume MSW modes are presented, with $k_{y,1}$s satisfying $n_{y,1}=k_{y,1}d/\pi\approx1, 2\ \mathrm{and}\ 3$, respectively. We note the characteristic numbers of $n_{y,1}$ are non-integral because of the boundary conditions. The dispersion surfaces show that volume MSWs with larger $n_{y,1}$s have higher frequencies with the same $k_x$ and $k_z$. Besides, all the surfaces appear saddle shapes with positive (negative) gradients along the $x (z)$ direction, implying that MSWs in the $x (z)$ direction is propagating forward (backward).  Additionally, the forward (backward) surface MSWs locate at the frequencies ranging from $\omega_{\mathrm{low}}$ to $\omega_{\mathrm{cut,\ f}}$ ($\omega_{\mathrm{cut,\ b}}$), while the volume MSW modes locate at the frequencies ranging from $\omega_H$ to $\omega_{\mathrm{low}}$. The increase of $d$ facilitates the merging of the multiple volume MSW modes and the increasing curvatures of the surfaces.

In principle, when the dispersion relation is determined, the ratios of $A_q\mathrm{s}$ and $\mathrm{\ }C_q\mathrm{s}$ to $C_1$ can be confirmed by evaluating Eqs. (\ref{BondCond}), given by
\begin{subequations}  \label{WaveFun}
\begin{align}
\label{WaveFuna}\frac{A_1}{C_1}&=\frac{\kappa k_x-k_\parallel\tanh{\left(k_\parallel l_2\right)}}{\left(\chi+1\right)k_{y,1}},\\
\label{WaveFunb}\frac{A_2}{C_1}&=\frac{1}{1+e^{2k_\parallel l_2}}=\frac{e^{-k_\parallel l_2}}{2\cosh{\left(k_\parallel l_2\right)}},\\
\label{WaveFunc}\frac{C_2}{C_1}&=\frac{1}{1+e^{-2k_\parallel l_2}}=\frac{e^{k_\parallel l_2}}{2\cosh{\left(k_\parallel l_2\right)}},\\
\label{WaveFund}\frac{A_0}{C_1}&=\frac{-\frac{A_1}{C_1}\sin{\left(k_{y,1}d\right)}+ \cos{\left(k_{y,1}d\right)}}{\left(e^{-2{k_\parallel l}_1}+1\right)e^{-k_\parallel d}},\\
\label{WaveFune}\frac{C_0}{C_1}&=\frac{-\frac{A_1}{C_1}\sin{\left(k_{y,1}d\right)}+ \cos{\left(k_{y,1}d\right)}}{\left(e^{2{k_\parallel l}_1}+1\right)e^{k_\parallel d}}.
\end{align}
\end{subequations}
Then wave functions describing the normalized intensity distributions of MSWs can be carried out by substituting Eqs. (\ref{WaveFun}) into (\ref{BondCond}). The analyses of wave functions will be presented in combination of the simulations for analyzing the performance of devices.

\section{NUMERICAL RESULTS AND ANALYSIS}\label{Simulations}
\subsection{Modeling and methods}\label{Model_metd}
To approach the performances of practical MSW devices, we conduct the finite-element electromagnetic (EM) simulations using Ansys High Frequency Structure Simulator (HFSS) software, in which the susceptibility tensors of saturated ferrites biased by $\mathbf{H}_\mathrm{0}$ are assigned by the definitions of Eqs. (\ref{Sus_Tensor}) and (\ref{Sus_Comp}) \cite{HFSSGuide}. Especially, the nearly zero in-plane demagnetizing factors imply that the static demagnetizing field induced by the magnetization can be neglected \cite{Yukawa1977}. Accordingly, the assignment of the magnetic biased source in HFSS simulation is exactly the internal direct current (DC) field equal to the $\mathbf{H}_\mathrm{0}$, without the need of a magneto-static solver like Maxwell3D, etc \cite{Marzall2021,Thalakkatukalathil2018}. Such simulations have also been demonstrated to be valid when neglecting the exchange interactions \cite{Robbins2022}.

\begin{figure}[htbp!]
  \centering
  \includegraphics[width=0.48\textwidth]{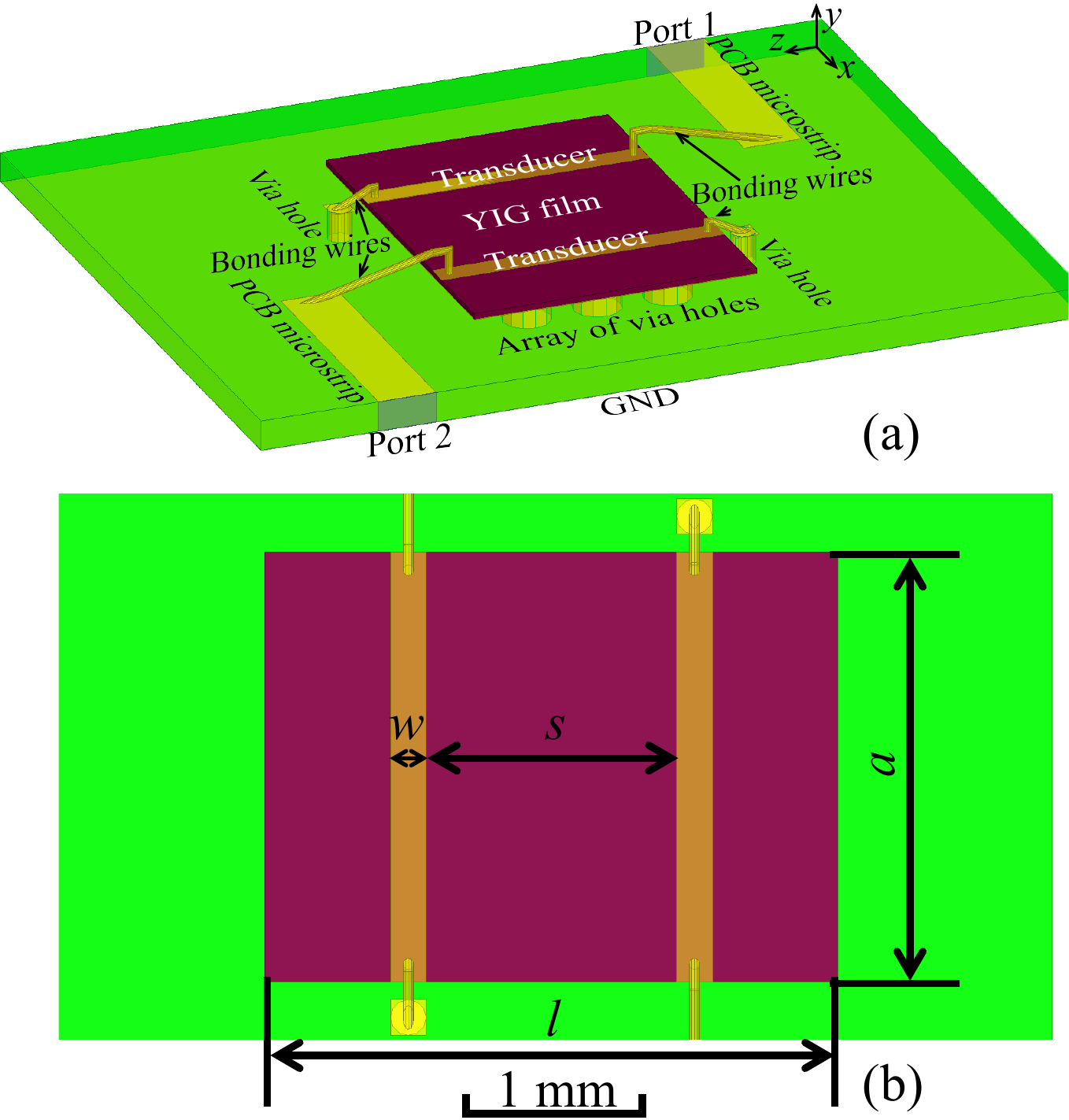}\\
  \caption{(a) The 3D and (b) the top views of the simulated model built by Ansys HFSS software. The thickness of PCB substrate, the width of PCB microstrip, the length and the width of the YIG film, the distance between the transducers and the width of the transducers are labelled as $h$, $b$, $l$, $a$, $s$, and $w$, respectively.} \label{fig4}
\end{figure}

The 3D and top views of the simulated model are shown in Fig. \ref{fig4}(a) and (b), respectively. MWs are either input or output from lumped Port 1 or Port 2 and guided by the printed circuit board (PCB) microstrips. Two transducers taking the form of microstrips are placed on top of the YIG film, whose bottom surface is metalized forming the FM structure, and grounded through the array of via holes contacting the GND layer of PCB. The transducers are connected to the PCB microstrips and the shorted terminals of via holes through bounding wires. Shorting the terminals creates the maximum current through the microstrips on the YIG film, and therefore the maximum excitation efficiency \cite{Stancil2012}. The dimensions are initially set as $d=50\ \mathrm{\mu m}$, $l=3.2\ \mathrm{mm}$, $a=2.4\ \mathrm{mm}$, $w=0.2\ \mathrm{mm}$ and $s=1.4\ \mathrm{mm}$, which can be parametrically simulated to evaluate their influence on the performances of devices. We consider the PCB substrate being FR4, whose relative permittivity $(\epsilon_\mathrm{r})$ is 4.2 \cite{CCWu2013}, which gives $h=0.254\ \mathrm{mm}$ and $b=0.49\ \mathrm{mm}$ for the $50\ \mathrm{\Omega}$ characteristic impedance matching. The conductivity of $4.1\times{10}^7\ \mathrm{S/m}$ are assigned on the transducers, the metallic layer at the bottom of the YIG film, the bonding wires and the PCB microstrips. The $\epsilon_\mathrm{r}$ of the YIG film is initially set to be 11.8 \cite{HZhao2004}.

The study of the scattering matrix parameters ($S$-parameters) of the system is a primary method for researching the MW properties of the MSW-based devices. Particularly, the forward and reverse transmission parameters $S_{21}$ and $S_{12}$ are essential for investigating the properties of MSWs \cite{Serga2010}.

\begin{figure}
  \centering
  \includegraphics[width=0.48\textwidth]{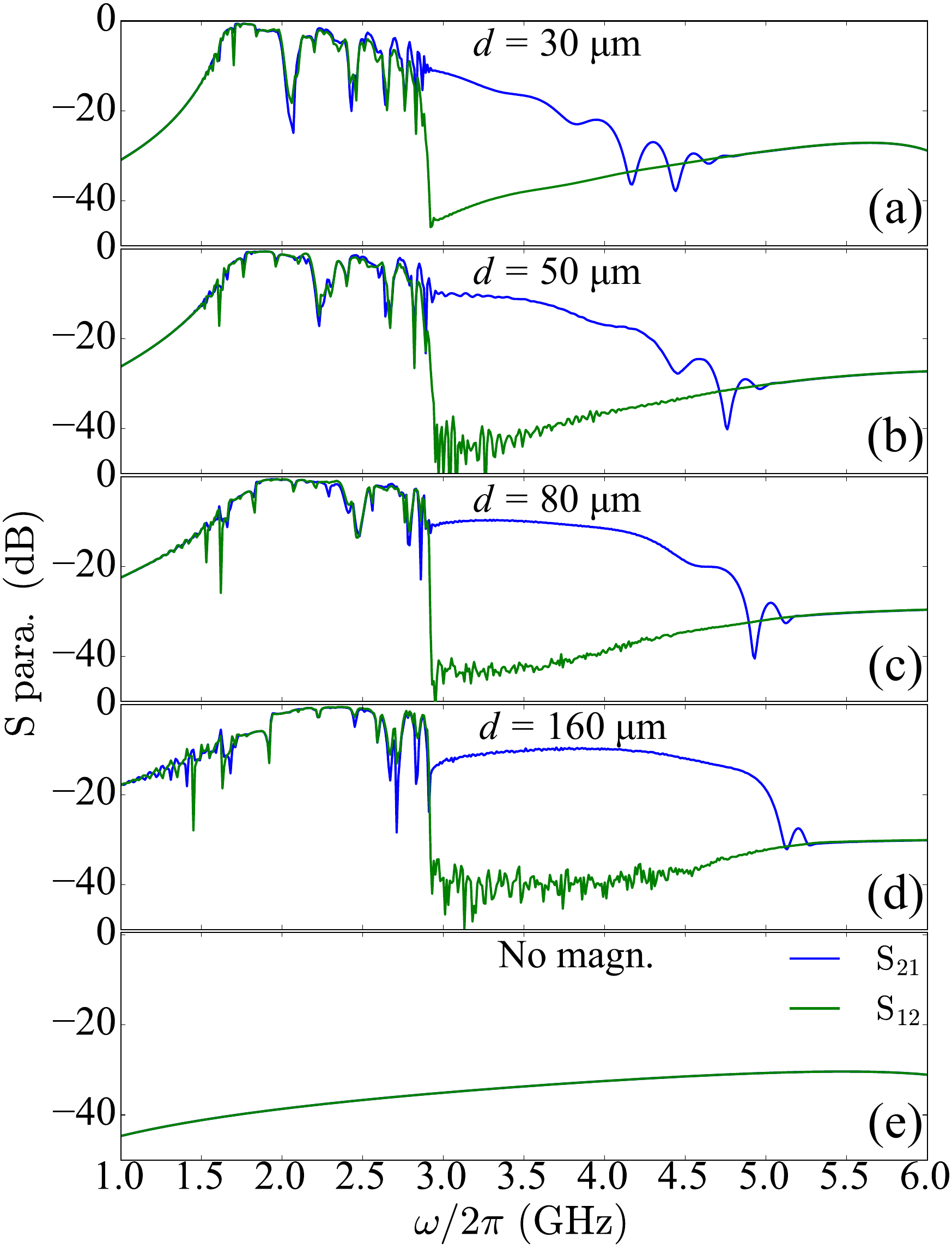}\\
  \caption{Simulated $S_{21}$ (blue curves) and $S_{12}$ (green curves) parameters with (a) $d=30\mathrm{\ \mu m}$, (b) $d=50\mathrm{\ \mu m}$, (c) $d=80\mathrm{\ \mu m}$, (d) $d=160\mathrm{\ \mu m}$ and (e) no magnetic film, respectively.} \label{fig5}
\end{figure}

The simulated transmission characteristics dependence on the ferrite thickness is surveyed with $d$ varied from 30 to 160 $\mathrm{{\mu} m}$, as plotted in Fig. \ref{fig5}(a)-(d). The transmission curves exhibit noteworthy nonreciprocities in the high frequency bands above $\omega_{\mathrm{cut,\ b}}$. Meanwhile, the $S_{21}$ curves in the high frequency bands present more flatness but higher ILs than those in the low frequency bands. Similar experimental results have been reported without theoretical analyses \cite{Geiler2022A,Geiler2022B}. The transmission curves with no magnetic films in the simulated model (Fig. \ref{fig4}) are plotted in Fig. \ref{fig5}(e). Through comparison, the multiple passbands presented in Fig. \ref{fig5}(a)-(d) are demonstrated to be caused by the MSWs propagating in the ferrite films. To verify this point, the normalized magnetic and electric fields are extracted for analyzing, with the representative magnitude distributions at the top surface of the ferrite film with $d=50\mathrm{\ \mu m}$ at $\omega/2\pi=1.9\ \mathrm{GHz}$ plotted in Fig. \ref{fig6}(a) and (b) (also see Supplemental movies Movie1.gif and Movie2.gif), respectively. Obviously, the energies are dominantly conveyed by the magnetic components of the fields, reconfirming that MSWs are magnetic waves.

\begin{figure}
  \centering
  \includegraphics[width=0.48\textwidth]{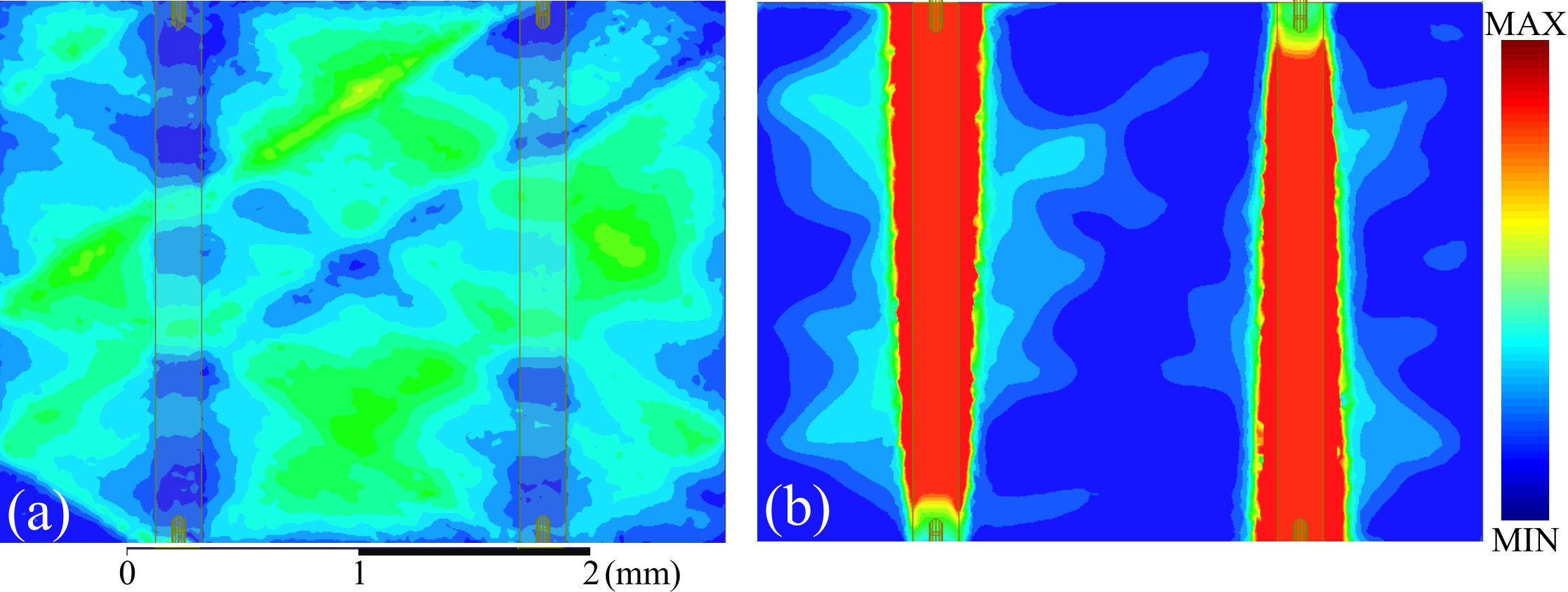}\\
  \caption{Normalized (a) magnetic and (b) electric field magnitude distributions at the top surface of the ferrite film with $d=50\mathrm{\ \mu m}$ at $\omega/2\pi=1.9\ \mathrm{GHz}$. The RF current is input from the left transducer (same hereinafter, if not stated otherwise). } \label{fig6}
\end{figure}

Below, we focus on the characteristics of the transmission curves induced by MSWs. The transmission curves ($S_{21}$ and $S_{12}$) present prominent nonreciprocities that $S_{21}$ is much larger than $S_{12}$ at the high frequency bands above $2.91\ \mathrm{GHz}$. It agrees well with the theoretical estimation that the upper frequency limits of backward and forward surface MSWs are $\omega_{\mathrm{cut,\ b}}$ and $\omega_{\mathrm{cut,\ f}}$, respectively. We refer to the frequency range between them as the high frequency bands. Correspondingly, the investigated low frequency bands range from $\omega_{\mathrm{low}}$ to $\omega_{\mathrm{cut,\ b}}$.

\subsection{High frequency characteristics}\label{HighFreqCh}

\begin{figure}
  \centering
  \includegraphics[width=0.48\textwidth]{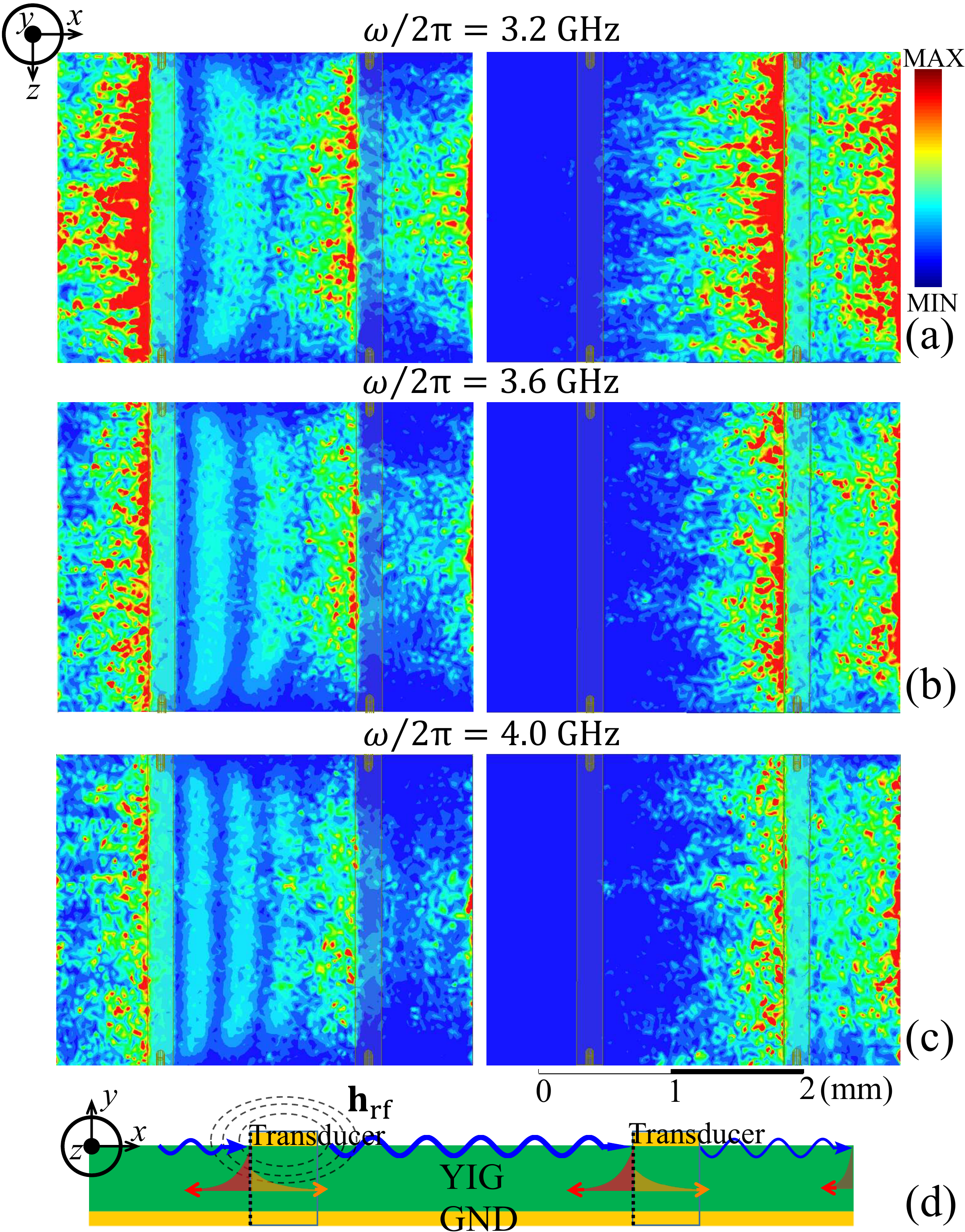}\\
  \caption{Normalized magnetic field magnitude distributions at the top surface of the ferrite film with $d=50\mathrm{\ \mu m}$ at (a) $3.2\ \mathrm{GHz}$, (b) $3.6\ \mathrm{GHz}$ and (c) $4.0\ \mathrm{GHz}$, respectively. The left (right) panels show the patterns with the MWs input from the left (right) transducers. (d) Schematic of the MSWs at high frequencies propagating in the simulated YIG films. The dashed vertical lines represent the FM/MFM boundaries. The two metallic transducers in combination with the grounded ferrite film constitute the MFM structures, indicated by the grey rectangles. The dashed ellipses represent the $\mathbf{h}_{\mathrm{rf}}$ induced by the input RF current. The wavy blue arrows represent the forward propagating MSWs, with the thickness indicating the intensities. The exponential red and yellow shapes with arrows represent the backward and forward evanescent MSWs at the FM/MFM boundaries, respectively. The exponential blue shape with arrow represents the backward evanescent MSWs totally reflected at the right edge of the ferrite film. } \label{fig7}
\end{figure}

In the high frequency bands, the $S_{21}$ curves appear distinct ILs, indicating substantial energy losses for the forward surface MSWs. For further analysis, the magnetic field magnitude distributions of the ferrite film are extracted for analysis, with the representative cases of $d=50\mathrm{\ \mu m}$ at $3.2\ \mathrm{GHz}$, $3.6\ \mathrm{GHz}$ and $4.0\ \mathrm{GHz}$ plotted in Fig. \ref{fig7}(a)-(c), respectively. When the MWs input from Port 1, the forward propagating MSWs between the transducers are observed to be perfectly plane waves, whose wave lengths are shortened with the $\omega$ increasing [see the left panels of  Fig. 7(a)-(c) and Supplemental movies Movie3-5.gif]. In addition, considerably intensive fields are induced on the left sides of the two transducers and the right edge of YIG films, meaning a significant amount of MSWs accumulated without being converted to MWs, thus leading to the high ILs shown in the $S_{21}$ curves. Similarly, the MSWs excited by the right transducers cannot propagate backward to the left transducers [see the right panels of  Fig. \ref{fig7}(a)-(c) and Supplemental movies Movie6-8.gif] due to the lack of backward modes according to the dispersion relations [see Fig. \ref{fig2}(b) and Fig. \ref{fig3}(b)], resulting in the even higher ILs presented in the $S_{12}$ curves and the consequent nonreciprocities.

Here, the transducers play two roles: (i) the converters for the RF currents and MSWs interconversion; (ii) the metallic layer on the MSWs' transmission path together with the grounded YIG film forming the MFM structure, as sketched in  Fig. 7(d). The $\mathbf{h}_{\mathrm{rf}}$ induced by the RF currents distribute around the input transducer. Therefore, the MSWs can be excited on both sides. However, the MSWs on the left side of the input transducer cannot propagate through the MFM structure, as has been discussed in Sec. \ref{Ana_cal} . Since no corresponding backward propagating modes exist in the dispersion relation [see the green curve in Fig. \ref{fig2}(b) and the viridian surface in Fig. \ref{fig3}(b)], most of the MSWs are reflected at the FM/MFM boundaries and accumulated and dissipated on the left side of both transducers. Meanwhile, a small portion of the MSWs can penetrate in an evanescent manner \cite{ZZhang2021,Verba2020}, pass through the MFM and keep on propagating in the FM. Finally, at the right edge of the YIG film, it would be totally reflected and decayed due to the forbiddance of backward propagating modes. The decay lengths of the backward waves are also shortened with the $\omega$ increasing, as illustrated in Fig. \ref{fig7}(a)-(c).

Moreover, the $S_{21}$ magnitude tends to decrease with increase of $\omega$ since the MSW excitation efficiencies decrease with increasing the $k_x$ \cite{Stancil2012,Ando2003,Sharawy1990,Lee1993,Weiss2023Theo}. Additionally, with the increase of $d$, the $S_{21}$ curve becomes flat with its magnitude slowly decreasing at high frequencies [see Fig. \ref{fig5}(a)-(d)]. This is attributed to the tiny variations of $k_x$s in the high frequency band [see the cyan curve in Fig. \ref{fig2}(b)]. We also note that the shapes of $S_{21}$ curves at high frequencies are related to the MSW excitation efficiencies, which are heavily determined by the current density distributions on the transducers. Under the uniform current distribution condition described by $J\left(x\right)=I/w$ with $I$ being the current, Fourier transformation gives the $k_x$ dependent efficiencies:
\begin{equation} \label{Eq31}
\begin{aligned}
j(k_x)=I\frac{\sin{\left(\frac{k_xw}{2}\right)}}{\frac{k_xw}{2}}.
\end{aligned}
\end{equation}
Hence, several minimum points emerge on the $S_{21}$ curves at some $k_x$s determined by  $j(k_x)=0$, one of which is $15.7\ \mathrm{rad/mm}$ by the present model parameters.  By comparing the simulated $S_{21}$ curves [see Fig. \ref{fig5}(a)-(d)] with the theoretical dispersion curves [see Fig. \ref{fig2}(b)], a good agreement between the theories and simulations is achieved.

\subsection{Low frequency characteristics}\label{LowFreqCh}

\begin{figure}
  \centering
  \includegraphics[width=0.40\textwidth]{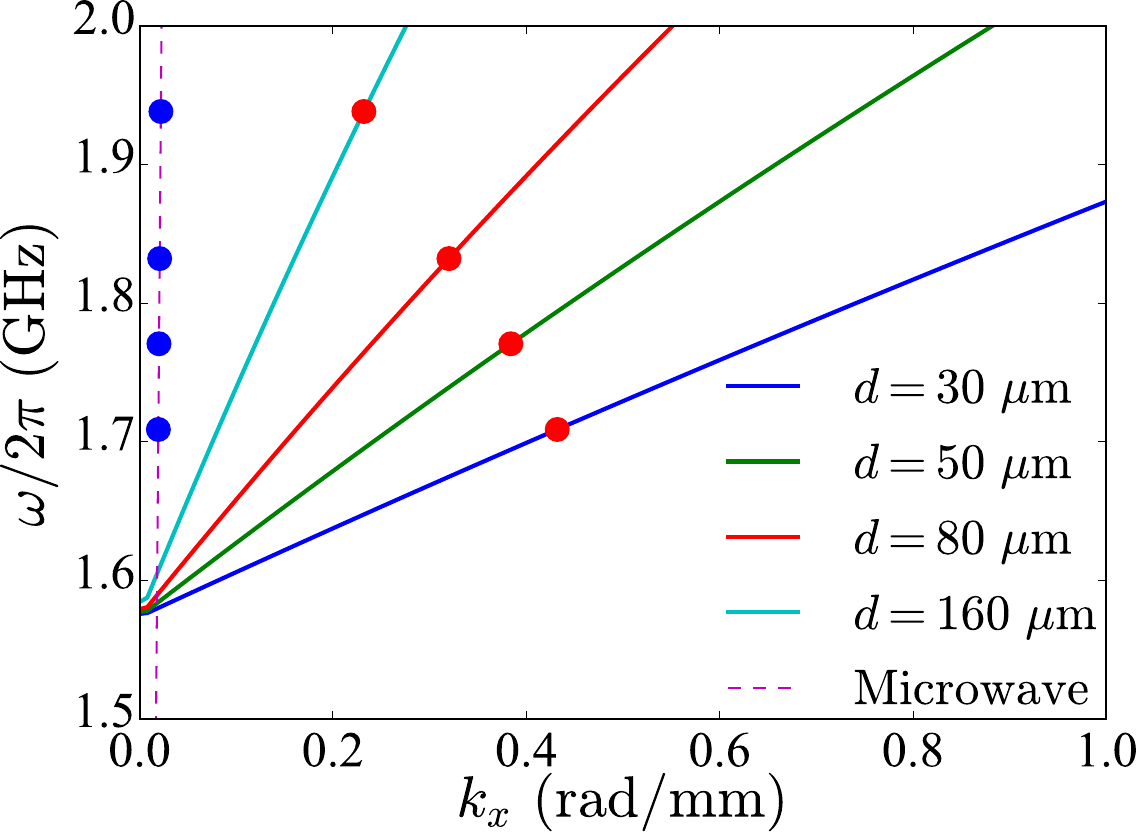}\\
  \caption{Zoomed in dispersion curves focused on the small wave vectors. Red and blue circles highlight the $k_x$s of MSWs and MWs at $\omega_{\mathrm{low}}$s, respectively.} \label{fig8}
\end{figure}

In the low frequency bands, it is firstly noticed that the low-frequency limit $\omega_{\mathrm{low}}/2\pi$ on efficient MSWs transmission does not match with the theoretical analysis, with the value being $1.57\ \mathrm{GHz}$ in the present model. The difference is mainly attributed to the magnetostatic approximation in Eq. (\ref{fig4}), which requires that $k_x$s of MSWs should be much larger than those of MWs at the same frequencies \cite{Stancil2012}.  In the zoomed in dispersion curves (Fig. \ref{fig8}), it is clear that the  $k_x$s of MSWs (red circles in Fig. \ref{fig8}) are at least one order of magnitude larger than those of MWs (blue circles in Fig. \ref{fig8}) at $\omega_{\mathrm{low}}$s. Furthermore, even though the $\omega_{\mathrm{low}}$ increases with $d$ increasing, the $k_x$ of MSW keeps decreasing, meaning that a thicker ferrite film can convey surface MSWs with longer decay lengths across the thickness, according to Eqs. (\ref{Eq29}) and (\ref{WaveFun}).

\begin{figure}
  \centering
  \includegraphics[width=0.48\textwidth]{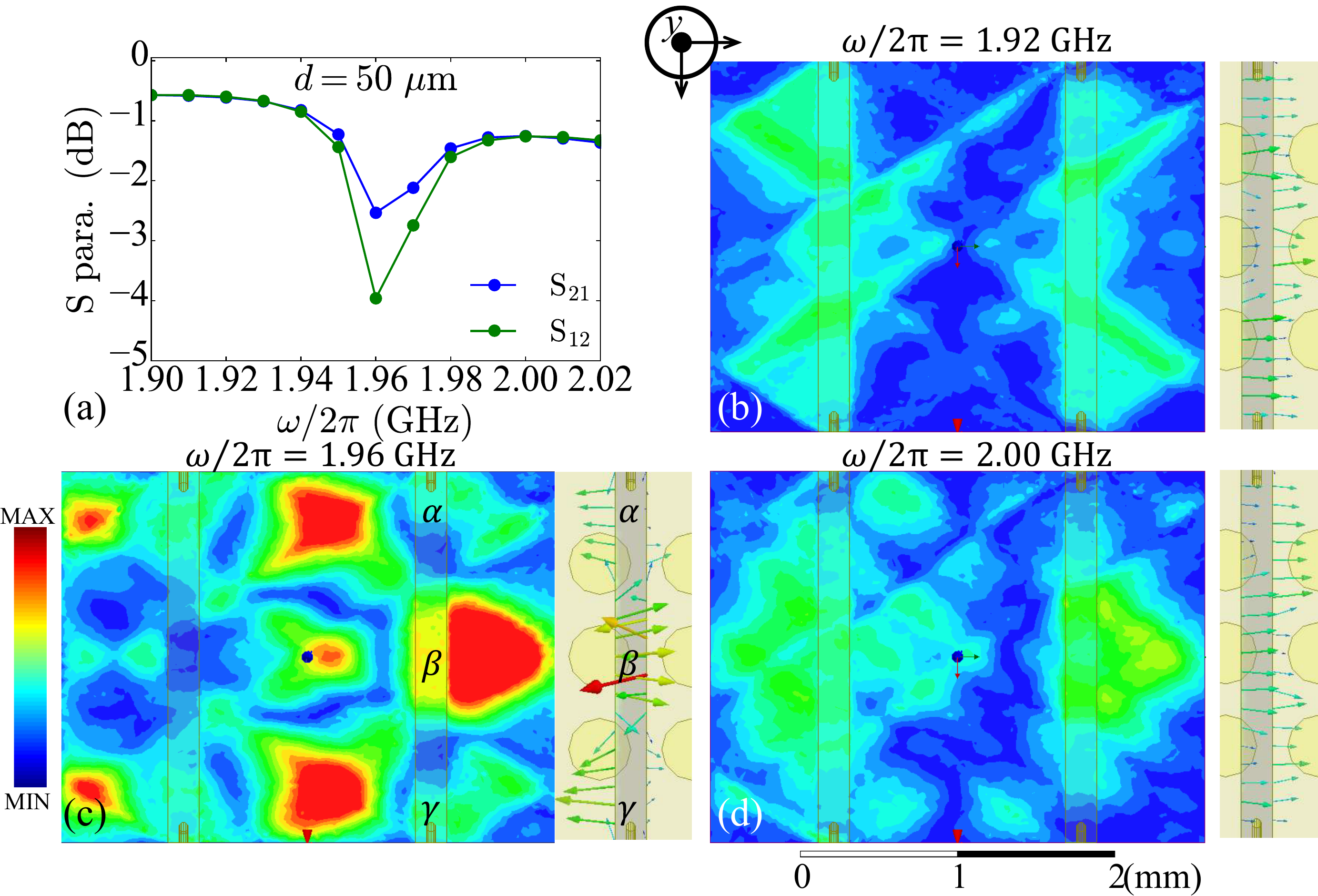}\\
  \caption{(a) Zoomed in simulated $S_{21}$ (blue) and $S_{12}$ (green) parameters focused on the first lowest dip at $1.96\ \mathrm{GHz}$. Normalized magnetic field magnitude distributions at the top surface of the ferrite film (left panels) and magnetic field vectors focusing on the receiving transducers (right panels) with $d=50\mathrm{\ \mu m}$ at (a) $1.92\ \mathrm{GHz}$, (b) $1.96\ \mathrm{GHz}$ and (c) $2.00\ \mathrm{GHz}$.} \label{fig9}
\end{figure}

\begin{figure}
  \centering
  \includegraphics[width=0.48\textwidth]{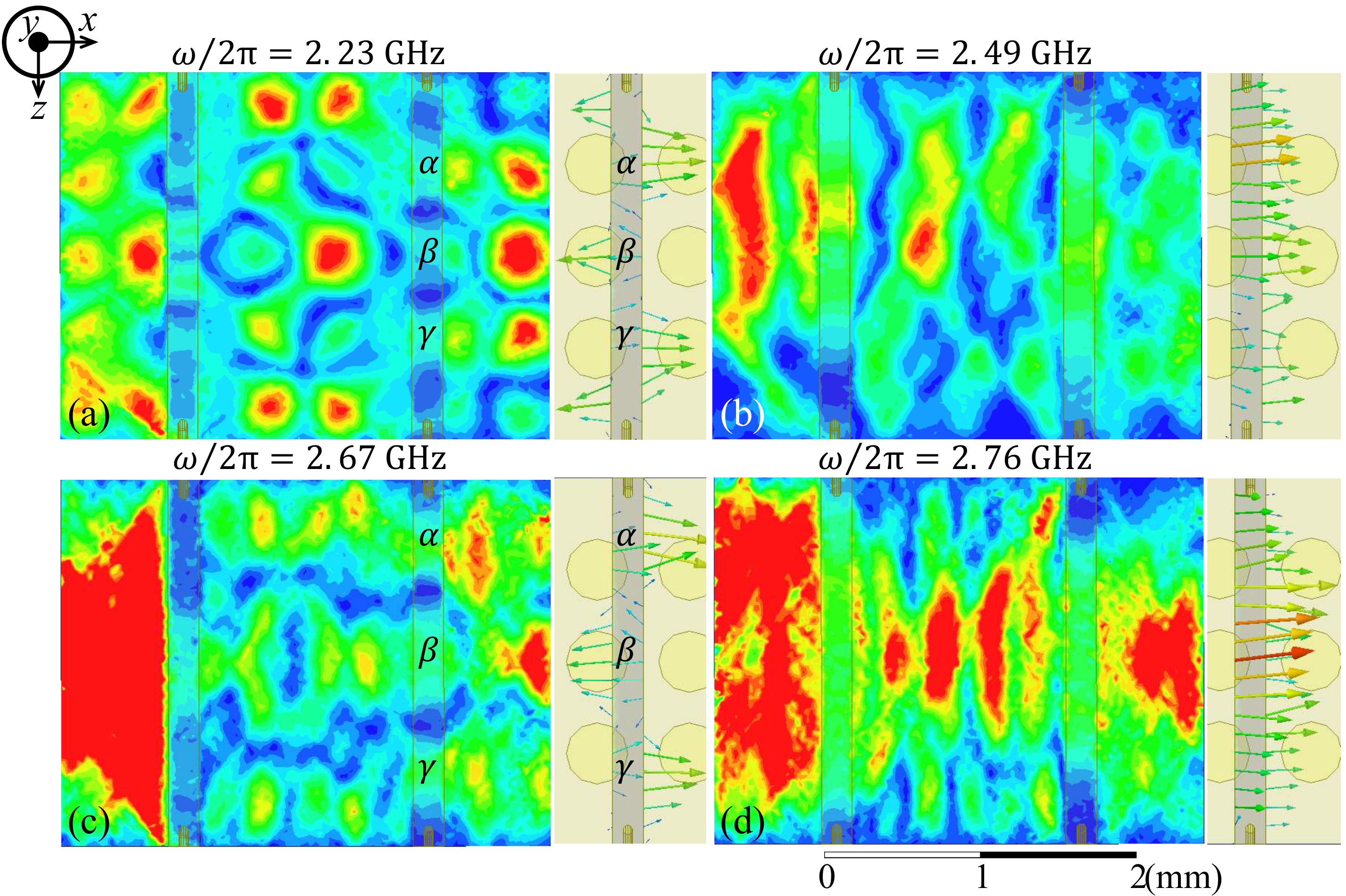}\\
  \caption{Normalized magnetic field magnitude distributions at the top surface of the ferrite film (left) and magnetic field vectors focusing on the receiving transducers (right) with $d=50\mathrm{\ \mu m}$ at (a) $2.23\ \mathrm{GHz}$, (b) $2.49\ \mathrm{GHz}$, (c) $2.67\ \mathrm{GHz}$ and (d) $2.76\ \mathrm{GHz}$.} \label{fig10}
\end{figure}

The other noticeable characteristics of the transmission curves at low frequency bands are the multiple dips, dividing the passbands into several sub-bands and resulting in the ripples. With the $d$ increasing, the dips drift toward higher frequencies and tend to merge at $\omega_H+0.5\omega_{M,1}$ [see Fig. \ref{fig5}(a)-(d)]. To gain a deeper insight, we extract the normalized magnetic field magnitude distributions for further analysis. The first lowest dip at $1.96\ \mathrm{GHz}$ in the passband of the transmission curve with $d=50\mathrm{\ \mu m}$ [see  Fig. \ref{fig5}(b) and Fig. \ref{fig9}(a)] is studied. The patterns of the MSWs propagating at $1.92\ \mathrm{GHz}$, $1.96\ \mathrm{GHz}$ and $2.00\ \mathrm{GHz}$ are plotted in  Fig. \ref{fig9}(b)-(d) (also see the corresponding animations in Supplemental movies Movie9-11.gif). It is observed that at the dip frequency [see Fig. \ref{fig9}(c)], higher width modes with $k_z=n\pi/a$ ($n$ is an odd number larger than one) are excited \cite{Kittel1958}, interfering with the main mode, which is similar with the self-focusing effect of the propagating spin waves in the micrometer-scale \cite{Demidov2008,Demidov2007,ZZhang2019}. Such interferences cause the splitting of the MSW beam into three portions [labelled as $\alpha$, $\beta$ and $\gamma$ in Fig. \ref{fig9}(c)]. However, the phase of $\beta$ portion takes almost $\pi$ difference with those of the $\alpha$ and $\gamma$ portions, as indicated by arrows in the right panel of Fig. \ref{fig9}(c) pointing at different directions (also see the Supplemental movie Movie12.gif). The phase differences result in the destructive superposition of the corresponding converted MWs and the consequent decreased efficiency of the ``MSWs to MWs" conversion at the receiving transducer on the right side.  At the frequencies away from the dips, like $1.92\ \mathrm{GHz}$ and $2.00\ \mathrm{GHz}$ [see Figs. \ref{fig9}(b) and (d)], the excitation of higher width modes is suppressed. The MSW beam arriving at the receiving transducer keeps almost the same phases along the entire width [see the right panels of Fig. \ref{fig9}(b) and (d) and the Supplemental movies Movie13.gif and Movie14.gif], rising the transmission efficiency. To verify the above analysis, we investigate the patterns at some other dips [see Fig. \ref{fig10}(a) and (c) and the Supplemental movies Movie15.gif and Movie16.gif] and those located between the dips [see Figs. \ref{fig10}(b) and (d) and the Supplemental movies Movie17.gif and Movie18.gif]. Apparently, higher width modes excited at the dips leads to the splitting of MSW beams into several portions [labelled as $\alpha$, $\beta$ and $\gamma$ in Fig. \ref{fig10}(a) and (c), respectively], but with quite different phases [see the right panels of Figs. \ref{fig10}(a) and (c) the Supplemental movies Movie19.gif and Movie20.gif]. The destructive superpositions are detrimental to the transmission efficiencies. On the contrary, at the frequencies away from the dips, the almost the same phases of the MSWs beams at the receiving transducers [see the right panels of Figs. \ref{fig10}(b) and (d) the Supplemental movies Movie21.gif and Movie22.gif] along the entire width favors decreasing the ILs. The same principles can be applied on analyzing the other transmission characteristics of the models with different $d$s, $a$s well as the other model parameters. To explore the excitation mechanisms of high-order spin waves, an even more complicated model considering the distribution of the surface current density on the microstrips, the boundary conditions and the wave functions of the magnetic fields in combination with the dispersion relations is required.

\section{EXPERIMENTAL VALIDATION}\label{ExpVald}

\subsection{Experimental preparation}\label{ExpPrep}

\begin{figure}
  \centering
  \includegraphics[width=0.48\textwidth]{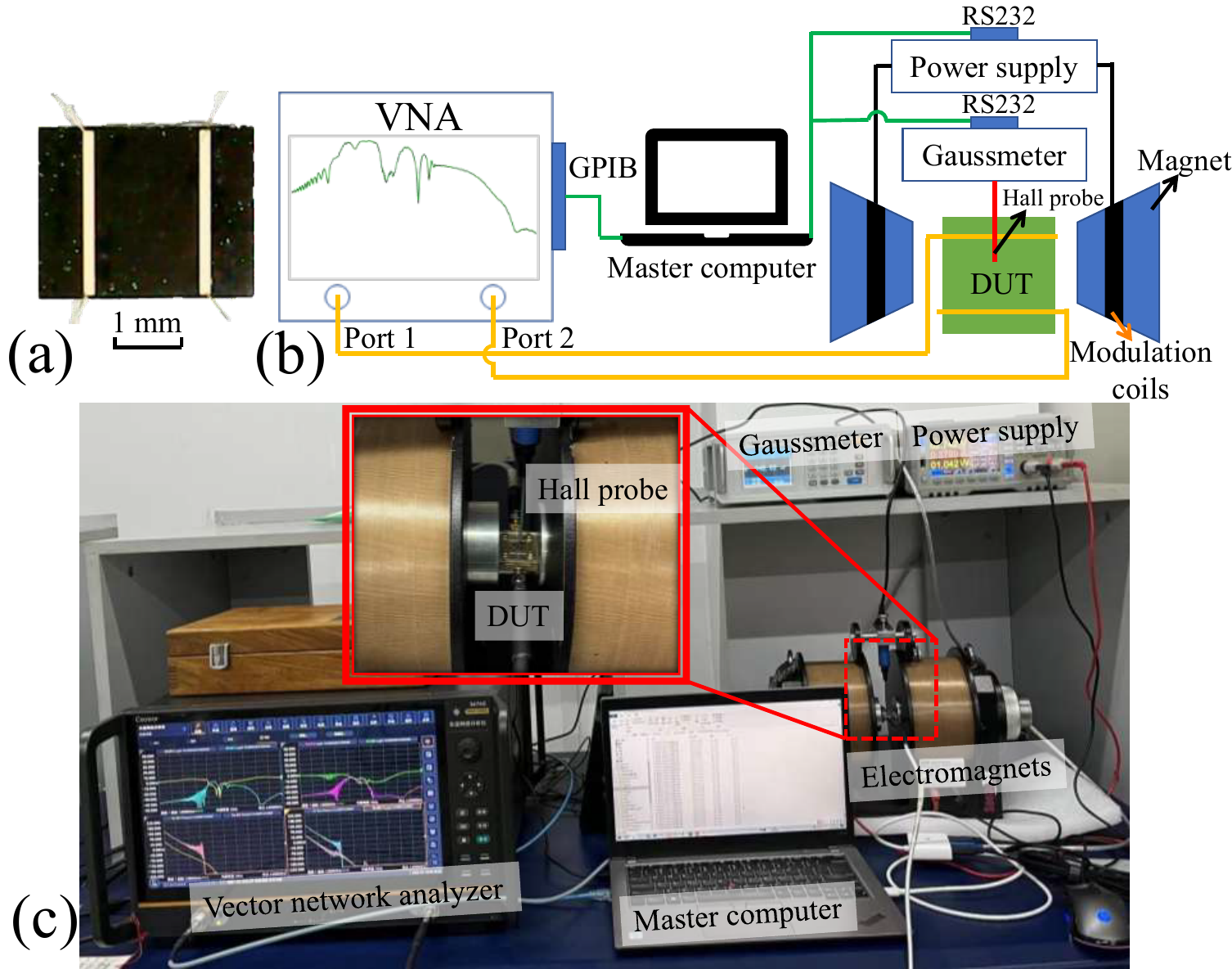}\\
  \caption{(a) Optical microscopic top view of the device. (b) Schematic diagram of the characterization of the DUT transmission curves. The yellow, green, and black lines represent the microwave, data, and DC current transmission lines, respectively. (c) Self-built experimental platform of DUT under various biased magnetic fields.} \label{fig11}
\end{figure}

To validate the above analysis experimentally, a proofing device under test (DUT) is fabricated according to the simulated model (see Fig. \ref{fig4}) using a commercially available YIG ferrite film, with the dimension being $3.2\ \mathrm{mm}\times2.4\ \mathrm{mm}\times50\ \mathrm{\mu m}$. The transducers are deposited on the film with the dimension of $0.2\ \mathrm{mm}\times2.4\ \mathrm{mm}\times1\ \mathrm{\mu m}$ and the distance of $s=1.4\ \mathrm{mm}$, as shown in Fig. \ref{fig11}(a). The transmission characteristics are measured using the self-built experimental platform, with the schematic diagram and the measurement setup shown in Figs. \ref{fig11}(b) and (c), respectively. The platform consists of a vector network analyzer (VNA), a gaussmeter, a direct current (DC) power supply and a master computer. The master computer is used for precisely tuning the $H_0$ by a feedback algorithm synchronously monitoring the gaussmeter and manipulating the DC power supply, and promptly reading the transmission data from the VNA. The input power is set to be $-20\ \mathrm{dBm}$ to avoid the nonlinear effects \cite{Venugopal2020}.

\subsection{Results and discussions}\label{ExpRslt}

\begin{figure}
  \centering
  \includegraphics[width=0.48\textwidth]{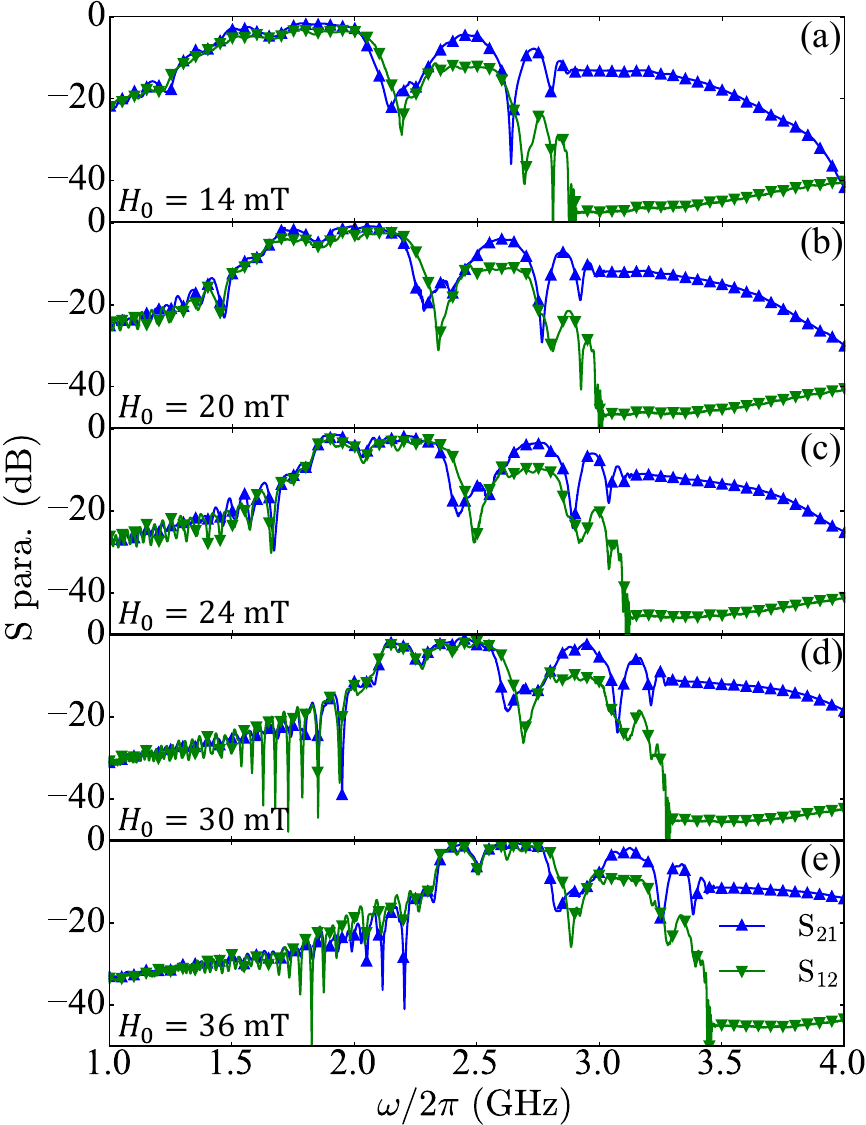}\\
  \caption{Measured $S_{21}$ (blue curves) and $S_{12}$ (green curves) parameters with (a) $H_0=14\mathrm{\ mT}$, (b)$H_0=20\mathrm{\ mT}$, (c) $H_0=24\mathrm{\ mT}$, (d) $H_0=30\mathrm{\ mT}$ and (e) $H_0=36\mathrm{\ mT}$, respectively.} \label{fig12}
\end{figure}

The transmission characteristics of the DUT are firstly demonstrated to be tunable by the $H_0$ (see Fig. \ref{fig12}), indicating that the passbands are caused by the propagations of MSWs in the ferrite film \cite{JWu2012,YZhang2020}. The DUT exhibits multiple passbands separated by some dips in the low frequency bands below $\omega_{\mathrm{cut,\ b}}$ for each $H_0$. Meanwhile, the ILs and the nonreciprocities becomes significant above $\omega_{\mathrm{cut,\ b}}$.  These features verify the analysis in Secs. \ref{Sec_Theo} and \ref{Simulations}. With $H_0$ increasing from $14\mathrm{\ mT}$ to $36\mathrm{\ mT}$, the minimal IL decreases from $1.6\ \mathrm{dB}$ to $0.5\ \mathrm{dB}$. The center frequency of the widest passband drifts from $1.85\ \mathrm{GHz}$ to $2.65\ \mathrm{GHz}$. The bandwidth of the widest passband decreases from $300\ \mathrm{MHz}$ to $210\ \mathrm{MHz}$. The measured performances also demonstrate that the proofing DUT can work as an effective magnetically tunable filter \cite{JWu2012,YZhang2020}.

\begin{figure}
  \centering
  \includegraphics[width=0.48\textwidth]{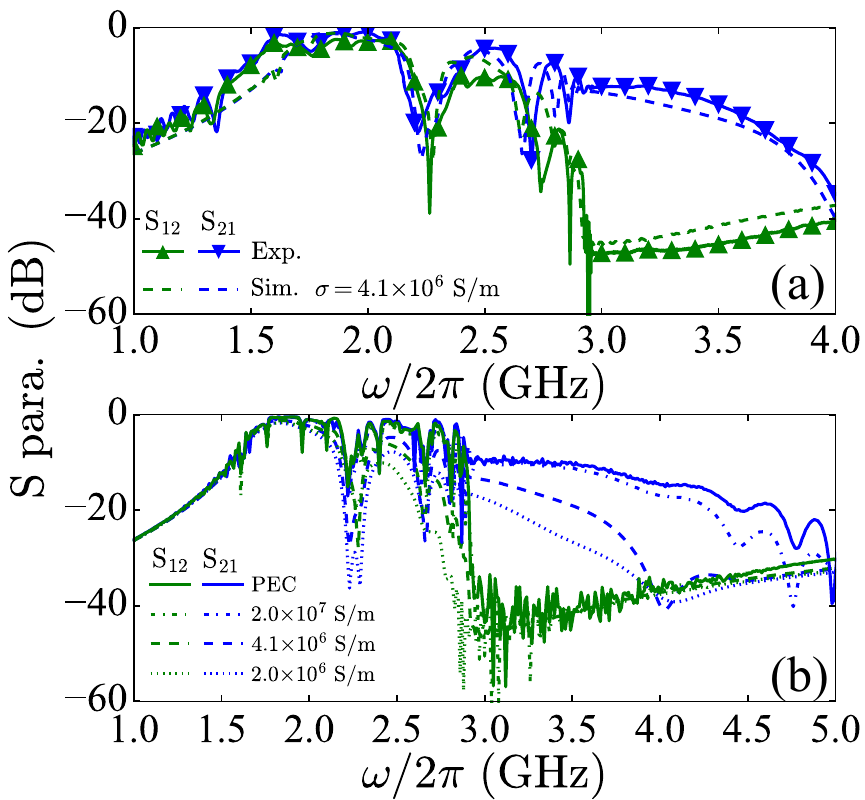}\\
  \caption{(a) Measured and simulated $S_{21}$ (blue curves) and $S_{12}$ (green curves) parameters of the DUT under $H_0=16.5\mathrm{\ mT}$. The simulation is performed with $\sigma=4.1\times{10}^6\ \mathrm{S/m}$. (b) Simulated $S_{21}$ (blue curves) and $S_{12}$ (green curves) parameters with the metallic layer at the bottom of YIG being PEC (solid lines), and the $\sigma=2.0\times{10}^7\ \mathrm{S/m}$ (dash-dot lines),  $\sigma=4.1\times{10}^6\ \mathrm{S/m}$ (dash-dot lines)and $\sigma=2.0\times{10}^6\ \mathrm{S/m}$ (dotted lines).} \label{fig13}
\end{figure}

Finally, the measured results are compared with the simulation results under $H_0=16.5\mathrm{\ mT}$. Limited by the advancement of the fabrication techniques, the conductivities of the metallic materials in the DUT are lower than those in the simulations performed in Sec \ref{Simulations}. Here, we find the simulation results with the conductivity of the metallic layer at the bottom of YIG ($\sigma$) being $4.1\times{10}^6\ \mathrm{S/m}$ agree well with the measured results, as shown in Fig. \ref{fig13} (a). Furthermore, we investigate the transmission characteristics dependence on the $\sigma$, with the simulated $S_{21}$ and $S_{12}$ curves obtained with various $\sigma$ plotted in  Fig. \ref{fig13} (b). Peculiarly, the ILs in the high frequencies reduce more than those in the low frequencies with the $\sigma$ increasing. Nevertheless, the ILs cannot reach to zero even with the perfect electric conductor (PEC) layer [see the solid blue line in Fig. \ref{fig13} (b)], since the MFM structure does not support the propagating surface MSWs as discussed in Sec. \ref{HighFreqCh}.

\section{CONCLUSIONS}\label{Concl}

Based on the above analyses and results, two possible methods can realize the MSW devices with broader passbands: (1) suppressing the excitation of higher width modes; (2) the strategies for surface MSWs passing through the MFM structures.  However, the higher width modes can be rather easily excited with the magnetic film taking the millimeter scale in the width and length directions and the micrometer scale in the thickness direction, since a lot of width modes merges in the dispersion relations (see Fig. \ref{fig3}). Possible strategies might be the deposition of high-quality YIG films with microscale dimensions directly on the metallic layer \cite{Aldosary2016,YSun2012} for the well separations of the multiple modes in the dispersion relations. On the other hand, the strategies of surface MSWs passing through the MFM structure can be difficult to realize in the present framework of a single ferrite layer. Investigations on the propagating MSWs in bilayers or even multilayers are necessary for the development of MSW devices with lower ILs in broader bands, in which several magnetic films should act as buffer layers in analog to the impedance matching techniques in MW engineering \cite{Pozar2021}.

In summary, we demonstrate that the characteristics of propagating MSWs play critical roles in the performance of the MSW devices.  We present a wave-propagation based analysis, according to which the characteristics of propagating MSWs in ferrite films with metallic structures can be predicted. We show that the high ILs in high-frequency bands are caused by the metallic nature of the transducers, while the dips and the ripples in low-frequency bands are caused by the complex interferences between the multiple width modes. Both simulation and experimental results verify the analyses with good agreements. Our findings shine a light on the MSWs propagating in ferrite films with metallic structures, helpful for designing MSW devices for being implanted in microwave systems.

\section{ACKNOWLEDGEMENTS}
\begin{acknowledgments}\label{Acknowledgments}
This work was supported by the National Natural Science Foundation of China (Grants No. 12404122, and No. 52202106) and  by Sichuan Science and Technology Program (Grant No. 2025ZNSFSC0868).
\end{acknowledgments}


\begin{thebibliography}{99}

\bibitem {Chumak2015} A. V. Chumak, V. I. Vasyuchka, A. A. Serga, and B. Hillebrands, Magnon spintronics, \href{https://www.nature.com/articles/nphys3347}{Nature Phys. \textbf{11}, 453 (2015)}.

\bibitem {Kruglyak2010} V. V. Kruglyak, S. O. Demokritov, and D. Grundler, Magnonics, \href{https://iopscience.iop.org/article/10.1088/0022-3727/43/26/264001/meta}{J. Phys. D: Appl. Phys. \textbf{43}, 264001 (2010)}.

\bibitem {Serga2010}A. A. Serga, A. V. Chumak, and B. Hillebrands, YIG magnonics \href{https://iopscience.iop.org/article/10.1088/0022-3727/43/26/264002/meta}{J. Phys. D: Appl. Phys. \textbf{43}, 264002 (2010)}.

\bibitem {Lenk2011}B. Lenk, H. Ulrichs, F. Garbs, and M. M{\"u}nzenberg, The building blocks of magnonics, \href{https://doi.org/10.1016/j.physrep.2011.06.003}{Phys. Rep. \textbf{507}, 107 (2011)}.

\bibitem {Owens1985}J. M. Owens, J. H. Collins, and R. L. Carter, System applications of magnetostatic wave devices, \href{https://link.springer.com/article/10.1007/BF01600088}{ Circuits Syst. Signal Process., \textbf{4}: 317 (1985)}.

\bibitem {Adam1991}J. D. Adam, M. R. Daniel, P. R. Emtage, and S. H. Talisa, in \href{https://doi.org/10.1016/B978-0-12-533015-2.50006-7}{Physics of Thin Films}, Vol. 15, edited by M.H. Francombe and J.L. Vossen (CA: Elsevier, 1991) Chap. 1, pp. 1{-}141.

\bibitem {Kasahara2017}K. Kasahara, M. Nakayama, X. Ya, K. Matsuyama, and T. Manago, Effect of distance between a magnet layer and an excitation antenna on the nonreciprocity of magnetostatic surface waves, \href{https://iopscience.iop.org/article/10.7567/JJAP.56.010309/meta}{Jpn. J. Appl. Phys. \textbf{56}, 010309 (2017)}.

\bibitem {Stancil2012}D. D. Stancil and A. Prabhakar, \textit{Spin Waves: Theory and Applications.} (Springer, New York, 2009)

\bibitem {Kabos2012}P. Kabos, and V. Stalmachov, \textit{Magnetostatic waves and their application.} (Springer Science and Business Media, 2012)

\bibitem {Damon1961}R. W. Damon, and J. R. Eshbach, Magnetostatic modes of a ferromagnet slab, \href{https://doi.org/10.1016/0022-3697(61)90041-5}{J. Phys. Chem. Solids \textbf{19}, 308 (1961)}.

\bibitem {Kalinikos1986}B. A. Kalinikos, and A. N. Slavin, Theory of dipole-exchange spin wave spectrum for ferromagnetic films with mixed exchange boundary conditions, \href{https://iopscience.iop.org/article/10.1088/0022-3719/19/35/014/meta}{J. Phys. C: Solid State Phys.  \textbf{19}, 7013 (1986)}.

\bibitem {JWu2012}J. Wu, X. Yang, S. Beguhn, J. Lou, and N. X. Sun, Nonreciprocal Tunable Low-Loss Bandpass Filters With Ultra-Wideband Isolation Based on Magnetostatic Surface Wave, \href{https://doi.org/10.1109/TMTT.2012.2222661}{IEEE Trans. Microw. Theory and Techn. \textbf{60}, 3959 (2012)}.

\bibitem {YZhang2020} Y. Zhang, D. Cai, C. Zhao, M. Zhu, Y. Gao, Y. Chen, X. Liang, H. Chen, J. Wang, Y. Wei, Y. He, C. Dong, N. Sun, M. Zaeimbashi, Y. Yang, H. Zhu, B. Zhang, K. Huang, and N. X. Sun, Nonreciprocal Isolating Bandpass Filter With Enhanced Isolation Using Metallized Ferrite, \href{https://doi.org/10.1109/TMTT.2020.3030784}{IEEE Trans. Microw. Theory and Techn. \textbf{68}, 5307 (2020)}.

\bibitem {XDu2024}X. Du, M. H. Idjadi, Y. Ding, T. Zhang, A. J. Geers, S. Yao, J. B. Pyo, F. Aflatouni, M. Allen, and R. H. Olsson, Frequency tunable magnetostatic wave filters with zero static power magnetic biasing circuitry, \href{https://doi.org/10.1038/s41467-024-47822-3}{Nat. Commun. \textbf{15}, 3582 (2024)}.

\bibitem {Adam2013}J. D. Adam, and F. Winter, Magnetostatic Wave Frequency Selective Limiters, \href{https://doi.org/10.1109/TMTT.2012.2222661}{IEEE Trans. Microw. Theory and Techn. \textbf{49}, 956 (2013)}.

\bibitem {Geiler2021} M. Geiler, S. Gillette, M. Shukla, P. Kulik, and A. l. Geiler, Microwave Magnetics and Considerations for Systems Design, \href{https://doi.org/10.1109/TMTT.2020.3030784}{IEEE J. Microwaves \textbf{1}, 438 (2021)}.

\bibitem {HLin2023}H. Lin, X. Shi, C. Dubs, M. Sanghadasa, and N. Sun, Compact and Passive Thin-Film Frequency-Selective Limiters, \href{https://doi.org/10.1038/s41467-024-47822-3}{IEEE Microw. Wireless Compon. Lett. \textbf{33}, 1155 (2023)}.

\bibitem {Adam2004}J. D. Adam, and S. N. Stitzer, MSW frequency selective limiters at UHF, \href{https://doi.org/10.1038/s41467-024-47822-3}{IEEE Trans. Magn. \textbf{40}, 2844 (2004)}.

\bibitem {Jamali2013}M. Jamali, J. H. Kwon, S.-M. Seo, K.-J. Lee, and H. Yang, Spin wave nonreciprocity for logic device applications, \href{https://doi.org/10.1038/srep03160}{Sci. Rep. \textbf{3}, 3160 (2013)}.

\bibitem {Klingler2015}S. Klingler, P. Pirro, T. Br?cher, B. Leven, B. Hillebrands, and A. V. Chumak, Spin-wave logic devices based on isotropic forward volume magnetostatic waves, \href{https://doi.org/10.1063/1.4921850}{Appl. Phys. Lett. \textbf{106}, 212406 (2015)}.

\bibitem {Kostylev2005}M. P. Kostylev, A. A. Serga, T. Schneider, B. Leven, and B. Hillebrands, Spin-wave logical gates, \href{https://doi.org/10.1063/1.2089147}{Appl. Phys. Lett. \textbf{87}, 153501 (2005)}.

\bibitem {Freer2019}B. Freer, and M. Geiler. Squint Reduction of L Band Phased Array Using Novel Miniature True Time Delay, \href{https://doi.org/10.1109/PAST43306.2019.9020773}{in Proc. IEEE Int. Symp. Phased Array Syst. Technol. (2019)}.

\bibitem {Vysotski2006}S. L. Vysotski\v{i}, G. T. Kazakov, A. V. Kozhevnikov, S. A. Nikitov, A. V. Romanov, and Y. A. Filimonov, Nondispersive delay line on magnetostatic waves, \href{https://doi.org/10.1134/S1063785006080098}{Tech. Phys. Lett. \textbf{32}, 3102 (2002)}.

\bibitem {Kobljanskyj2002}Y. V. Kobljanskyj, G. A. Melkov, V. M. Pan, V. S. Tiberkevich, and A. N. Slavin, Active magnetostatic wave delay line for microwave signals, \href{https://doi.org/10.1109/TMAG.2002.802481}{IEEE Trans. Magn. \textbf{38}, 3102 (2002)}.

\bibitem {Bajpai1988}S. N. Bajpai, R. L. Carter, and J. M. Owens, Insertion loss of magnetostatic surface wave delay lines, \href{https://doi.org/10.1109/22.3492}{IEEE Trans. Microw. Theory and Techn. \textbf{36}, 132 (1988)}.

\bibitem {Wahi1982}P. Wahi, and Z. Turski, Magnetostatic Wave Dispersive Delay Line, \href{https://doi.org/10.1109/22.3492}{IEEE Trans. Microw. Theory and Techn. \textbf{30}, 2031 (1982)}.

\bibitem {Bajpai1989}S. N. Bajpai, Insertion loss of electronically variable magnetostatic wave delay lines, \href{https://doi.org/10.1038/s41598-020-69364-6}{IEEE Trans. Microw. Theory and Techn. \textbf{37}, 1529 (1989)}.

\bibitem {Daniel1985}M. R. Daniel, J. D. Adam, and P. R. Emtage, Dispersive delay at gigahertz frequencies using magnetostatic waves,  \href{https://doi.org/10.1007/BF01600076}{Circ., Syst. Signal Pr. \textbf{4}, 115 (1985)}.

\bibitem {Szulc2020}K. Szulc, P. Graczyk, M. Mruczkiewicz, G. Gubbiotti, and M. Krawczyk, Spin-Wave Diode and Circulator Based on Unidirectional Coupling, \href{https://doi.org/10.1103/PhysRevApplied.14.034063}{Phys. Rev. Applied \textbf{14}, 034063 (2020)}.

\bibitem {Seewald2010}C. K. Seewald, and J. R. Bray, Ferrite-Filled Antisymmetrically Biased Rectangular Waveguide Isolator Using Magnetostatic Surface Wave Modes, \href{https://doi.org/10.1109/TMTT.2010.2047919}{IEEE Trans. Microw. Theory and Techn. \textbf{58}, 1493 (2021)}.

\bibitem {Shichi2015}S. Shichi, N. Kanazawa, K. Matsuda, S. Okajima, T. Hasegawa, T. Okada, T. Goto, H. Takagi, and M. Inoue, Spin wave isolator based on frequency displacement nonreciprocity in ferromagnetic bilayer, \href{https://doi.org/10.1063/1.4915101}{ J. Appl. Phys. \textbf{117}, 17D125 (2015)}.

\bibitem {Thiancourt2024}G.Y. Thiancourt, S.M. Ngom, N. Bardou, and T. Devolder, Unidirectional spin waves measured using propagating-spin-wave spectroscopy, \href{https://doi.org/10.1103/PhysRevApplied.22.034040}{ Phys. Rev. Applied \textbf{22}, 034040 (2024)}.

\bibitem {Ganguly1975}A. K. Ganguly, and D. C. Webb, Microstrip Excitation of Magnetostatic Surface Waves: Theory and Experiment, \href{https://doi.org/10.1109/TMTT.1975.1128733}{IEEE Trans. Microw. Theory and Techn. \textbf{23}, 998 (1975)}.

\bibitem {Ganguly1978}A. K. Ganguly, D. C. Webb, and C. Banks, Complex Radiation Impedance of Microstrip-Excited Magnetostatic-Interface Waves, \href{https://doi.org/10.1109/TMTT.1975.1128733}{IEEE Trans. Microw. Theory and Techn. \textbf{26}, 444 (1978)}.

\bibitem {Emtage1978}P. R. Emtage, Interaction of magnetostatic waves with a current, \href{https://doi.org/10.1063/1.325452}{J. Appl. Phys. \textbf{49}, 4475 (1978)}.

\bibitem {Sushruth2020}M. Sushruth, M. Grassi , K. Ait-Oukaci, D. Stoeffler, Y. Henry , D. Lacour, M. Hehn,U. Bhaskar , M. Bailleul, T. Devolder, and J.-P. Adam, Electrical spectroscopy of forward volume spin waves in perpendicularly magnetized materials, \href{https://doi.org/10.1103/PhysRevResearch.2.043203}{ Phys. Rev. Research \textbf{2}, 043203  (2020)}.


\bibitem {YZhang2018} Y. Zhang, T. Yu, J. Chen, Y. Zhang, J. Feng, S. Tu, H. Yu, Antenna design for propagating spin wave spectroscopy in ferromagnetic thin films, \href{http://dx.doi.org/10.1016/j.jmmm.2017.04.048}{  J. Magn. Magn. Mater. \textbf{450}, 24 (2018)}.

\bibitem {Ando1998} Y. Ando, N. Guan, K. i. Yashiro, and S. Ohkawa, Excitation of magnetostatic surface wave by coplanar waveguide transducers, \href{https://doi.org/10.1109/APMC.1997.659294}{IEICE Trans. Electron. \textbf{81}, 1942 (1998)}.

\bibitem {Birt2012}D. R. Birt, K. An, M. Tsoi, S. Tamaru, D. Ricketts, K. L. Wong, P. Khalili Amiri, K. L. Wang, and X. Li, Deviation from exponential decay for spin waves excited with a coplanar waveguide antenna, \href{https://doi.org/10.1063/1.4772798}{Appl. Phys. Lett. \textbf{101}, 187601 (2012)}.

\bibitem {Celegato2015}F. Celegato, M. Co\"{i}sson, O. Khan, M. Kuepferling, A. Magni, C. Ragusa, A. Rahim, C. Portesi, and W. Wang, Comprehensive Theoretical and Experimental Analysis of Spin Waves in Magnetic Thin Film, \href{https://doi.org/10.1109/TMAG.2014.2360317}{IEEE Trans. Magn. \textbf{51}, 1 (2015)}.

\bibitem {Nakrap2006}I. A. Nakrap, A. N. Savin, and Y. P. Sharaevskii, Effect of magnetized ferromagnetic film on electrodynamic characteristics of a meander microstrip line, \href{https://link.springer.com/article/10.1134/S1064226906030077}{J. Commun. Technol. Electron. \textbf{51}, 303 (2006)}.

\bibitem {Collins1973}J. H. Collins, and F. A. Pizzarello, Propagating magnetic waves in thick films A complementary technology to surface wave acoustics,  \href{https://doi.org/10.1080/00207217308938446}{Int. J. Electron. \textbf{34}, 319 (1973)}.

\bibitem {Sethares1979}J. C. Sethares, Magnetostatic Surface-Wave Transducers,  \href{https://doi.org/10.1109/TMTT.1979.1129760}{IEEE Trans. Microw. Theory and Techn. \textbf{27}, 902 (1979)}.

\bibitem {Chang2014}H. Chang, P. Li, W. Zhang, T. Liu, A. Hoffmann, L. Deng, and M. Wu, Nanometer-Thick Yttrium Iron Garnet Films With Extremely Low Damping,  \href{https://doi.org/10.1021/acsami.7b19684}{IEEE Magn. Lett. \textbf{5}, 1 (2014)}.

\bibitem {Liu2014}T. Liu, H. Chang, V. Vlaminck, Y. Sun, M. Kabatek, A. Hoffmann, L. Deng, and M. Wu, Ferromagnetic resonance of sputtered yttrium iron garnet nanometer films, \href{https://doi.org/10.1063/1.4852135}{J. Appl. Phys.  \textbf{115}, 17A501 (2014)}.

\bibitem {Ding2020}J. Ding, T. Liu, H. Chang, and M. Wu, Sputtering Growth of Low-Damping Yttrium-Iron-Garnet Thin Films, \href{https://doi.org/10.1109/LMAG.2020.2989687}{IEEE Magn. Lett. \textbf{11}, 1 (2020)}.

\bibitem {Freire2003} M. J. Freire, R. Marques, and F. Medina, Full-wave analysis of the excitation of magnetostatic-surface waves by a semi-infinite microstrip transducer - theory and experiment, \href{https://doi.org/10.1109/TMTT.2003.808633}{IEEE Trans. Microw. Theory and Techn. \textbf{51}, 903 (2003)}.

\bibitem {Emtage1982}P. R. Emtage, Generation of magnetostatic surface waves by a microstrip, \href{https://doi.org/10.1063/1.331346}{J. Appl. Phys. \textbf{53}, 5122 (1982)}.


\bibitem {Aguilera2004}J. Aguilera, M. J. Freire, R. Marques, and F. Medina, Quasi-TEM model of magnetostatic-surface wave excitation in microstrip lines, \href{https://doi.org/10.1109/LMWC.2004.837067}{IEEE Microw. Wireless Compon. Lett. \textbf{53}, 5122 (1982)}.

\bibitem {Kalinikos1981}B. A. Kalinikos, Spectrum and linear excitation of spin waves in ferromagnetic films, \href{https://doi.org/10.1007/BF00941342}{Sov. Phys. J. \textbf{24}, 718 (1981)}.

\bibitem {Ando2003}Y. Ando, G. Ning, K. Yashiro, S. Ohkawa, and M. Hayakawa, An analysis of excitation of magnetostatic surface waves in an in-plane magnetized YIG film by the integral kernel expansion method, \href{https://doi.org/10.1109/TMTT.2002.807827}{IEEE Trans. Microw. Theory and Techn. \textbf{51}, 492 (2003)}.

\bibitem {Freire2003_IL}M. J. Freire, R. Marques, and F. Medina, Insertion loss of magnetostatic-surface wave transducers --transmission-line model and experiment, \href{https://doi.org/10.1109/TMTT.2003.817440}{IEEE Trans. Microw. Theory and Techn. \textbf{51}, 2126 (2003)}.

\bibitem {Sharawy1990}E. B. El-Sharawy, and R. W. Jackson, Full-wave analysis of an infinitely long magnetic surface wave transducer, \href{https://doi.org/10.1109/22.130967}{IEEE Trans. Microw. Theory and Techn. \textbf{38}, 730 (1990)}.

\bibitem {Lee1993}J.-H. Lee, and J.-W. Ra, Full-wave calculation of the radiation impedance of microstrip-excited magnetic surface waves, \href{https://doi.org/10.1002/mop.4650060715}{Microw. Opt. Technol. Lett. \textbf{6}, 441 (1993)}.

\bibitem {Weiss2023Theo}C. Weiss, M. Bailleul, and M. Kostylev, Excitation and reception of magnetostatic surface spin waves in thin conducting ferromagnetic films by coplanar microwave antennas. Part I: Theory, \href{https://doi.org/10.1016/j.jmmm.2022.170103}{J. Magn. Magn. Mater. \textbf{565}, 170103 (2023)}.

\bibitem {Weiss2023Exp} C. Weiss, M. Grassi, Y. Roussign\'{e}, A. Stashkevich, T. Schefer, J. Robert, M. Bailleul, and M. Kostylev, Excitation and reception of magnetostatic surface spin waves in thin conducting ferromagnetic films by coplanar microwave antennas. Part II: Experiment, \href{https://doi.org/10.1016/j.jmmm.2022.170002}{J. Magn. Magn. Mater. \textbf{565}, 170002 (2023)}.

\bibitem {Sethares1985}J. C. Sethares, and I. J. Weinberg, Theory of MSW transducers, \href{https://doi.org/10.1007/BF01600072}{Circ., Syst. Signal Pr. \textbf{4}, 41 (1985)}.

\bibitem {Chakrabarti1999}S. Chakrabarti, and D. Bhattacharya, Magnetostatic volume waves in lossy YIG film backed by a metal of finite conductivity, \href{https://doi.org/10.1109/22.775448}{IEEE Trans. Microw. Theory and Techn. \textbf{47}, 1132 (1999)}.

\bibitem {YLi1987}L. Yi, M. Koshiba, and M. Suzuki, Finite-Element Solution of Planar Inhomogeneous Waveguides for Magnetostatic Waves, \href{https://doi.org/10.1109/TMTT.1987.1133739}{IEEE Trans. Microw. Theory and Techn. \textbf{35}, 731 (1987)}.

\bibitem {Vyzulin1993}S. A. Vyzulin, A. E. Rosenson, and S. A. Shekh, The magnetostatic waves in ferrite film with losses, \href{https://doi.org/10.1109/22.238528}{IEEE Trans. Microw. Theory and Techn. \textbf{41}, 1070 (1993)}.

\bibitem {Tsutsumi1996}M. Tsutsumi, T. Fukusako, and S. Yoshida, Propagation characteristics of the magnetostatic surface wave in the YBCO-YIG film-layered structure, \href{https://doi.org/10.1109/22.536023}{IEEE Trans. Microw. Theory and Techn. \textbf{44}, 1410 (1996)}.

\bibitem {Masuda1974}M. Masuda, N. S. Chang, and Y. Matsuo, Magnetostatic Surface Waves in Ferrite Slab Adjacent to Semiconductor (Short Papers), \href{https://doi.org/10.1109/TMTT.1974.1128185}{IEEE Trans. Microw. Theory and Techn. \textbf{22}, 132 (1974)}.

\bibitem {Tsutsumi1976}M. Tsutsumi, T. Bhattacharyya, and N. Kumagai, Effect of the Magnetic Perturbation on Magnetostatic Surface-Wave Propagation,\href{https://doi.org/10.1109/TMTT.1976.1128913}{IEEE Trans. Microw. Theory and Techn. \textbf{24}, 591 (1976)}.

\bibitem {Geiler2022A}M. Geiler, FSL having a free standing YIG film, U. S. Patents. 11,349,185 (2022)

\bibitem {Geiler2022B}M. Geiler, Frequency Selective Limiter Having an Enhanced Bandwidth, U. S. Patents, 17/804,151 (2022).

\bibitem {Abo2013}G. S. Abo, Y. Hong, J. Park, J. Lee, W. Lee, and B. Choi, Definition of Magnetic Exchange Length, \href{https://doi.org/10.1109/TMAG.2013.2258028}{IEEE Trans. Magn. \textbf{49}, 4937 (2013)}.

\bibitem {Polder1949}D. Polder, VIII. On the theory of ferromagnetic resonance,  \href{https://doi.org/10.1080/14786444908561215}{Philos. Mag. \textbf{40}, 99 (1949)}.

\bibitem {Yukawa1977}T. Yukawa, J.-i. Yamada, K. Abe, and J.-i. Ikenoue, Effects of Metal on the Dispersion Relation of Magnetostatic Surface Waves, \href{https://iopscience.iop.org/article/10.1143/JJAP.16.2187/meta}{Jpn. J. Appl. Phys. \textbf{16}, 2187 (1977)}.

\bibitem {HFSSGuide}\textit{HFSS10 help guide, available Online}.

\bibitem {Yukawa1977}A. Aharoni, Demagnetizing factors for rectangular ferromagnetic prisms, \href{https://doi.org/10.1063/1.367113}{J. Appl. Phys. \textbf{83}, 3432 (1998)}.

\bibitem {Marzall2021}L. Marzall, D. Psychogiou, and Z. Popovi\'{c}, Microstrip Ferrite Circulator Design With Control of Magnetization Distribution, \href{https://doi.org/10.1109/TMTT.2020.3045995}{IEEE Trans. Microw. Theory and Techn. \textbf{69}, 1217 (2021)}.

\bibitem {Thalakkatukalathil2018}V. V K Thalakkatukalathil, A. Chevalier, V. Laur, G. Verissimo, P. Queffelec, L. Qassym, and R. Lebourgeois, Electromagnetic modeling of anisotropic ferrites¡ªApplication to microstrip Y-junction circulator design, \href{https://doi.org/10.1063/1.5026482}{J. Appl. Phys. \textbf{123},  234503 (2018)}.

\bibitem {Robbins2022}M. Robbins, D. Connelly, and J. Chisum, Modeling Thick Metal in Forward Volume Spin Wave Transducers, \href{https://doi.org/10.1063/1.367113}{IEEE Microw. Wireless Compon. Lett. \textbf{32}, 684 (2022)}.

\bibitem {CCWu2013}C.-C. Wu, and C.-F. Yang, Measuring the frequency-dependent dielectric properties of microwave composites using simple measurement methods,  \href{https://doi.org/10.1016/j.jallcom.2013.07.044}{J. Alloys Compd. \textbf{581}, 636 (2013)}.

\bibitem {HZhao2004}H. Zhao, J. Zhou, Y. Bai, Z. Gui, and L. Li, Effect of Bi-substitution on the dielectric properties of polycrystalline yttrium iron garnet, \href{https://doi.org/10.1016/j.jmmm.2004.03.014}{J. Magn. Magn. Mater. \textbf{280}, 208 (2004)}.

\bibitem {ZZhang2021}Z. Zhang, H. Yang, Z. Wang, Y. Cao, and P. Yan, Strong coupling of quantized spin waves in ferromagnetic bilayers, \href{https://doi.org/10.1103/PhysRevB.103.104420}{Phys. Rev. B \textbf{103}, 104420 (2021)}.

\bibitem {Verba2020}R. Verba, V. Tiberkevich, and A. Slavin, Spin-wave transmission through an internal boundary: Beyond the scalar approximation, \href{https://doi.org/10.1103/PhysRevB.101.144430}{Phys. Rev. B \textbf{101}, 144430 (2020)}.

\bibitem {Kittel1958}C. Kittel, Excitation of Spin Waves in a Ferromagnet by a Uniform rf Field, \href{https://doi.org/10.1103/PhysRev.110.1295}{Phys. Rev. \textbf{110}, 1295 (1958)}.

\bibitem {Demidov2008}V. E. Demidov, S. O. Demokritov, K. Rott, P. Krzysteczko, and G. Reiss, Mode interference and periodic self-focusing of spin waves in permalloy microstripes, \href{https://doi.org/10.1103/PhysRevB.77.064406}{Phys. Rev. B \textbf{77}, 064406 (2008)}.

\bibitem {Demidov2007}R. Verba, V. Tiberkevich, and A. Slavin, Spin-wave transmission through an internal boundary: Beyond the scalar approximation, \href{https://doi.org/10.1063/1.2825421}{Appl. Phys. Lett. \textbf{91}, 252504 (2007)}.

\bibitem {ZZhang2019}Z. Zhang, M. Vogel, J. Holanda, J. Ding, M. B. Jungfleisch, Y. Li, J. E. Pearson, R. Divan, W. Zhang, A. Hoffmann, Y. Nie, and V. Novosad, Controlled interconversion of quantized spin wave modes via local magnetic fields, \href{https://doi.org/10.1103/PhysRevB.100.014429}{Phys. Rev. B \textbf{100}, 014429 (2019)}.

\bibitem {Venugopal2020}A. Venugopal, T. Qu, and R. H. Victora, Nonlinear Parallel-Pumped FMR: Three and Four Magnon Processes, \href{https://doi.org/10.1109/TMTT.2019.2952128}{IEEE Trans. Microw. Theory and Techn. \textbf{68}, 602 (2020)}.

\bibitem {Aldosary2016}M. Aldosary, J. Li, C. Tang, Y. Xu, J.-G. Zheng, K. N. Bozhilov, and J. Shi, Platinum/yttrium iron garnet inverted structures for spin current transport, \href{https://doi.org/10.1063/1.4953454}{Appl. Phys. Lett. \textbf{108}, 242401 (2016)}.

\bibitem {YSun2012}Y. Sun, Y.-Y. Song, and M. Wu, Growth and ferromagnetic resonance of yttrium iron garnet thin films on metals, \href{https://doi.org/10.1063/1.4747465}{Appl. Phys. Lett. \textbf{101}, 082405 (2012)}.

\bibitem {Pozar2021}D. M. Pozar, \textit{Microwave engineering: theory and techniques} (John wiley and sons, 2021).
\end{thebibliography}
\end{document}